\PassOptionsToPackage{unicode}{hyperref}
\PassOptionsToPackage{hyphens}{url}
\documentclass[
  12pt,
]{article}
\usepackage{lmodern}
\usepackage{setspace}
\usepackage{amssymb,amsmath}
\usepackage{ifxetex,ifluatex}
\ifnum 0\ifxetex 1\fi\ifluatex 1\fi=0 
  \usepackage[T1]{fontenc}
  \usepackage[utf8]{inputenc}
  \usepackage{textcomp} 
\else 
  \usepackage{unicode-math}
  \defaultfontfeatures{Scale=MatchLowercase}
  \defaultfontfeatures[\rmfamily]{Ligatures=TeX,Scale=1}
\fi
\IfFileExists{upquote.sty}{\usepackage{upquote}}{}
\IfFileExists{microtype.sty}{
  \usepackage[]{microtype}
  \UseMicrotypeSet[protrusion]{basicmath} 
}{}
\makeatletter
\@ifundefined{KOMAClassName}{
  \IfFileExists{parskip.sty}{%
    \usepackage{parskip}
  }{
    \setlength{\parindent}{0pt}
    \setlength{\parskip}{6pt plus 2pt minus 1pt}}
}{
  \KOMAoptions{parskip=half}}
\makeatother
\usepackage{xcolor}
\IfFileExists{xurl.sty}{\usepackage{xurl}}{} 
\IfFileExists{bookmark.sty}{\usepackage{bookmark}}{\usepackage{hyperref}}
\hypersetup{
  pdftitle={Contextualizing E-values for Interpretable Sensitivity to Unmeasured Confounding Analyses},
  pdfauthor={Lucy D'Agostino McGowan, PhD*{[}1{]}, Robert Greevy, Jr., PhD{[}2{]}; 1. Department of Mathematics and Statistics, Wake Forest University, Winston-Salem NC USA 2. Department of Biostatistics, Vanderbilt University, Nashville TN, USA; *Corresponding author: lucydagostino@gmail.com},
  hidelinks,
  pdfcreator={LaTeX via pandoc}}
\usepackage[margin=1in]{geometry}
\usepackage{longtable,booktabs}
\usepackage{etoolbox}
\makeatletter
\patchcmd\longtable{\par}{\if@noskipsec\mbox{}\fi\par}{}{}
\makeatother
\IfFileExists{footnotehyper.sty}{\usepackage{footnotehyper}}{\usepackage{footnote}}
\makesavenoteenv{longtable}
\usepackage{graphicx}
\makeatletter
\def\maxwidth{\ifdim\Gin@nat@width>\linewidth\linewidth\else\Gin@nat@width\fi}
\def\maxheight{\ifdim\Gin@nat@height>\textheight\textheight\else\Gin@nat@height\fi}
\makeatother
\setkeys{Gin}{width=\maxwidth,height=\maxheight,keepaspectratio}
\makeatletter
\def\fps@figure{htbp}
\makeatother
\providecommand{\tightlist}{%
  \setlength{\itemsep}{0pt}\setlength{\parskip}{0pt}}
\newlength{\cslhangindent}
\newenvironment{cslreferences}%
  {\setlength{\parindent}{0pt}%
  \everypar{\setlength{\hangindent}{\cslhangindent}}\ignorespaces}%
  {\par}

\title{Contextualizing E-values for Interpretable Sensitivity to Unmeasured Confounding Analyses}
\author{Lucy D'Agostino McGowan, PhD*{[}1{]}, Robert Greevy, Jr., PhD{[}2{]} \and 1. Department of Mathematics and Statistics, Wake Forest University, Winston-Salem NC USA 2. Department of Biostatistics, Vanderbilt University, Nashville TN, USA \and *Corresponding author: \href{mailto:lucydagostino@gmail.com}{\nolinkurl{lucydagostino@gmail.com}}}
\date{}

\begin{document}
\maketitle

\setstretch{2}
\newpage

\hypertarget{abstract}{%
\section{Abstract}\label{abstract}}

The strength of evidence provided by epidemiological and observational studies is inherently limited by the potential for unmeasured confounding. Researchers should present a quantified sensitivity to unmeasured confounding analysis that is contextualized by the study's observed covariates. VanderWeele and Ding's E-value provides an easily calculated metric for the magnitude of the hypothetical unmeasured confounding required to render the study's result inconclusive. We propose the \emph{Observed Covariate E-value} to contextualize the sensitivity analysis' hypothetical E-value within the actual impact of observed covariates, individually or within groups. We introduce a sensitivity analysis figure that presents the Observed Covariate E-values, on the E-value scale, next to their corresponding \emph{observed bias effects}, on the original scale of the study results. This \emph{observed bias plot} allows easy comparison of the hypothetical E-values, Observed Covariate E-values, and observed bias effects. We illustrate the methods with a specific example and provide a supplemental appendix with modifiable code that teaches how to implement the method and create a publication quality figure.

\hypertarget{introduction}{%
\section{Introduction}\label{introduction}}

The strength of the evidence provided by observational studies is inherently limited by the potential influence of unmeasured confounding variables. This limitation should neither be ignored nor used as a blanket dismissal of all observational studies' findings. Every observational study with a statistically significant finding should include a quantified sensitivity to unmeasured confounding analysis, a type of quantitative bias analysis (Lash et al. 2014; Lash, Fox, and Fink 2011). However, a 2008 systematic review by Groenwold et al.~showed such analyses were rarely done (Groenwold et al. 2008). They examined 174 observational studies in five general medical journals and five epidemiological journals published between January 2004 and April 2007. While the potential for unobserved confounding was reported in 102 (58.6\%) of reviewed articles, 15 (8.6\%) commented on the potential effect of such remaining confounding and only 4 (2.3\%) conducted a sensitivity analysis to estimate the potential impact of unobserved confounding. To see if the landscape had improved since then, we performed a review of 90 observational studies with statistically significant findings published in 2015 in the Journal of the American Medical Association, the New England Journal of Medicine, and the American Journal of Epidemiology. We saw little improvement with 41 (45.6\%) mentioning the issue of unmeasured confounding as a limitation and only 4 (4.4\%) performing a quantitative sensitivity analysis. Even when sensitivity analyses are performed, they can remain difficult for clinically oriented readers to understand. These deficiencies reveal the need for simple and interpretable quantified sensitivity to unmeasured confounding analyses.

One challenge of translating these methods into common practice is finding the right level of simplification. Consider the rule of thumb, ``In a study with binary outcomes and binary exposures the relative risk may be off by a factor of 2, but unlikely to be off more than that.'' (Van Belle 2011). Under simplified assumptions, this rule applies widely to generalized regression settings yielding relative risks, odds ratios, and hazard ratios. However, the criteria may be too liberal for studies missing one or more variables known to be strong confounders and too conservative for studies that adjust for all major known confounders. It ignores the study design's quality. A good sensitivity analysis should guide the reader through evaluating the quality of covariate adjustment that has been done and the potential for unmeasured confounding sufficient to change the study's conclusions. A well designed study that has controlled for several important confounders via matching, weighting, and/or regression-based covariate adjustment can provide the context in which the hypothetical unmeasured confounder's properties should be viewed.

Researchers need a sensitivity analysis that is simple enough to be widely implemented and contextualized enough to allow for useful interpretation.

VanderWeele and Ding recently suggested a tipping point sensitivity analysis simplification called the E-value (Ding and VanderWeele 2016; VanderWeele and Ding 2017). Here the tipping point is the E-value that corresponds to unmeasured confounding just strong enough to render the study results inconclusive at a 5\% significance level. The E-value offers a simple, quantified sensitivity analysis; however, without additional context, readers will have difficulty judging whether the tipping point E-value is large or small for a given study. We introduce the \emph{Observed Covariate E-value}, which quantifies the impact of each observed covariate on the E-value scale. These Observed Covariate E-values place the sensitivity analysis' E-value into the study context, grounding a seemingly abstract measure in what was observed.

\hypertarget{background}{%
\section{Background}\label{background}}

The main objective of a tipping point sensitivity analysis is to report the qualities of an unmeasured confounder needed to change the statistical significance of one's findings. For example, a hazard ratio of 1.25 with a 95\% confidence interval (1.1, 1.5) would no longer be significant at the \(\alpha = .05\) level if adjusting for a hypothetical unmeasured confounder shifted the interval to include 1.0. The ``tipping point'' analysis would find the weakest hypothetical confounder that did this, i.e.~that shifted the interval's lower bound to 1.0.

To determine whether an exposure, \(Z\), is associated with an outcome, \(Y\), one can observe whether the risk ratio, odds ratio, or hazard ratio of \(Z\) is equal to 1. As a tipping point analysis, we are interested in which values of an unmeasured confounder would cause the lower or upper confidence interval of the association measure to cross the null; we refer to this bound closest to the null as the ``limiting bound'', or \(LB\). These tipping point analysis extend from a large body of research on sensitivity analyses for unmeasured confounding (Cornfield et al. 1959; Bross 1966; Schlesselman 1978; Rosenbaum and Rubin 1983; Lin, Psaty, and Kronmal 1998; Greenland 1998, 2001, 2003, 2005; Robins, Rotnitzky, and Scharfstein 2000; Brumback et al. 2004; Schneeweiss 2006; McCandless, Gustafson, and Levy 2007, 2008; VanderWeele, Hernán, and Robins 2008; VanderWeele 2008b, 2008a, 2013; VanderWeele and Arah 2011; VanderWeele, Mukherjee, and Chen 2012; Hsu and Small 2013).

Researchers have been developing methods focusing on sensitivity analyses for unmeasured confounding for decades. In 1959, it was well known that there existed an association between smoking and lung cancer, but debate raged as to whether that was a causal relationship. Cornfield et al.~engaged in a discussion about the association between smoking and lung cancer (Cornfield et al. 1959). They derived the association between smoking and lung cancer in the event that this association was due solely to a binary unmeasured confounder. In this capacity, Cornfield quantified the prevalence of a binary unmeasured confounder in the exposed and unexposed population that would be necessary to fully nullify the observed association between smoking and lung cancer. Cornfield demonstrated ``if cigarette smokers have 9 times the risk of nonsmokers for developing lung cancer, and this is not because cigarette smoke is a causal agent, but only because cigarette smokers produce hormone X, then the proportion of hormone-X-producers among cigarette smokers must be at least 9 times greater than that of nonsmokers'' (Cornfield et al. 1959). In 1966, Bross coined the ``Size Rule'' (Bross 1966). Similar to Cornfield et al., Bross described the impact of a single unmeasured confounder on a given unadjusted effect by estimating what the risk ratio of the exposure effect would be if there was really no exposure effect. In 1978, Schlesselman allowed the association between the exposure and outcome to vary (Schlesselman 1978). In 1983, Rosenbaum and Rubin moved the conversation forward by allowing categorical covariate adjustment for the exposure-outcome effect (Rosenbaum and Rubin 1983). In 1998, Lin Psaty, and Kronmal generalized the advancement of Rosenbaum and Rubin by framing the sensitivity analysis within a regression framework (Lin, Psaty, and Kronmal 1998). They show that the observed association between \(Z\) and \(Y\) can be adjusted based on the size and prevalence of an independent unmeasured confounder \(U\), for a binary unmeasured confounder, and the size and mean difference between exposure groups for a continuous unmeasured confounder. Under the assumption that the sensitivity parameters are fixed, the variance of the observed effect is the same as the variance of the adjusted effect. This allows all adjustments to apply to confidence intervals the same way they would apply to point estimates. Lin et al.~algebraically derive equations to update biased estimates in log-linear regression for unmeasured confounders. Simulations show that these sensitivity analyses can be extended to the logistic regression and censored survival time cases under most conditions (Lin, Psaty, and Kronmal 1998). The relationship for the binary unmeasured confounder is as shown in Equation \eqref{eq:tip1}.

\begin{equation}
LB_{adj}=LB_{obs}\frac{RR_{UD} p_0+(1-p_0)}{RR_{UD} p_1+(1-p_1)}
\label{eq:tip1}
\end{equation}

Where \(LB_{adj}\) is the limiting bound of the risk ratio, odds ratio, or hazard ratio for \(Z\) adjusting for the unmeasured confounder and known confounders, \(LB_{obs}\) is the observed limiting bound obtained from the model including known confounders but excluding the unmeasured confounder, \(p_1\) and \(p_0\) are the prevalences of the unmeasured confounder in the exposed and unexposed populations, respectively, and \(RR_{UD}\) is the association between the unmeasured confounder and the outcome both in the presence and absence of the exposure, i.e.~with the assumption of no interaction. Notice here the unmeasured confounder is assumed to be binary, as we are estimating prevalences in the exposed and unexposed populations. Using a similar equation, Lin et al.~derive the relationship between a continuous unmeasured confounder (normally distributed, \(U\sim N(\mu_Z, 1)\)) and an outcome.

Building on this methodology, Ding and VanderWeele offer an ``assumption free'' method that no longer requires that the unmeasured confounder be binary, but rather represents this relationship as a risk ratio, in the binary case represented as \(RR_{EU} = p_1/p_0\) (Ding and VanderWeele 2016). They further recommend reporting the minimum \(RR_{UD}\) needed to tip under a particular \(RR_{EU}\). In the binary case, this is equivalent to setting \(p_1\) to 1, varying \(p_0\) from 0 to 1. Since we are interested in the tipping point such that the original association is no longer statistically significant, the adjusted limiting bound, \(LB_{adj}\) is set equal to 1 (Equation \eqref{eq:tip10}).

\begin{equation}
1 = LB_{obs}\frac{RR_{UD}/RR_{EU}+(1-1/RR_{EU})}{RR_{UD}}
\label{eq:tip10}
\end{equation}

While this method is described as ``assumption free'', it is important to highlight how it is solved in the binary case, that is the prevalence in the exposed group is assumed to be 1. This may represent an unrealistic extreme. This led to the introduction of the E-value.

\hypertarget{e-value}{%
\section{E-value}\label{e-value}}

VanderWeele and Ding further suggest focusing on the point that minimizes the strength of association, on the risk ratio scale, that an unmeasured confounder would need to have with both the exposure and outcome, conditional on the measured covariates, to explain away an observed exposure-outcome association (Ding and VanderWeele 2016; VanderWeele and Ding 2017). They call this value an ``E-value'' (Equation \eqref{eq:evalue}).

\begin{equation}
\textrm{E-value} = LB_{obs} + \sqrt{LB_{obs}\times(LB_{obs}-1)}
\label{eq:evalue}
\end{equation}

This E-value demonstrates the joint minimum strength of association with both the exposure and outcome needed to tip the analysis (allow the limiting bound to cross one, i.e.~\(LB_{adj} = 1\)).

Here, the limiting bound, \(LB_{obs}\) is assumed to be greater than one. If the limiting bound is less than one, it can be replaced with \(LB_{obs}^* = 1 / LB_{obs}\).

Similar to the equations above, the E-value is built for a risk ratio, i.e.~\(LB_{obs}\) is a risk ratio. VanderWeele and Ding specify that when the outcome is relatively rare the E-value in Equation \eqref{eq:evalue} can be used directly with an odds ratio or hazard ratio. For common outcomes, they recommend calculating an approximate E-value by taking the square root of the odds ratio, e.g.~for the limiting bound \(LB_{obs, OR}\):

\begin{equation}
\textrm{E-value} \approx \sqrt{LB_{obs, OR}} + \sqrt{\sqrt{LB_{obs, OR}}\times(\sqrt{LB_{obs, OR}}-1)}
\label{eq:evalue-or}
\end{equation}

Similarly, for a hazard ratio, \(LB_{obs, HR}\), with a common outcome, they recommend plugging the following limiting bound into Equation \eqref{eq:evalue}: \(LB_{obs}\approx\left(1-0.5^{\sqrt{(LB_{obs,HR})}}\right)/\left(1-0.5^{\sqrt{(1/LB_{obs,HR})}}\right)\).

This quantity, the E-value, is ideal in its simplicity in that each study can have a single number summary. In simplifying the relationship, however, there are some assumptions that need to be made. For example, since the E-value is an extension of Ding and VanderWeele's ``assumption free'' sensitivity analysis as shown in Equation \eqref{eq:tip10}, if the E-value were describing a binary confounder, the prevalence in the exposed group is assumed to be 1. This may be an extreme circumstance that would be unlikely to occur, making it difficult to determine the likelihood of missing a confounder that would tip the given analysis.

\hypertarget{observed-covariate-e-value}{%
\section{Observed Covariate E-value}\label{observed-covariate-e-value}}

Many of the existing methods attempt to quantify the unmeasured confounding needed to ``tip'' a given analysis, but do not provide context for how likely it is that an unmeasured confounder of such a magnitude exists. Similarly, there are methods for attempting to estimate the magnitude of a known unmeasured confounder to assess the impact this confounder would have on the given analysis, but this requires having a specific confounder in mind. We propose a hybrid approach that uses the E-value to quantify the unmeasured confounding needed to ``tip'' the analysis, and subsequently ``grounds'' this value in the observed covariates to give some context to help understand the type of confounder that could have the ``tipping'' impact on a given study. Using the same methodology as VanderWeele and Ding, we propose the Observed Covariate E-value. If one were interested in a different tipping point, we could reintroduce \(LB_{adj}\) from Equation \label{eq:tip1} into the equation as shown in Equation \eqref{eq:evalueadj}.

\begin{equation}
\textrm{E-value}_{OC} = \frac{LB_{obs}}{LB_{adj}} + \sqrt{\frac{LB_{obs}}{LB_{adj}}\times\left(\frac{LB_{obs}}{LB_{adj}}-1\right)}
\label{eq:evalueadj}
\end{equation}

The above equation assumes \(LB_{obs}\) is greater than \(LB_{adj}\). If \(LB_{obs}\) is less than \(LB_{adj}\), the equation changes to the following.

\begin{equation}
\textrm{E-value}_{OC} = \frac{LB_{adj}}{LB_{obs}} + \sqrt{\frac{LB_{adj}}{LB_{obs}}\times\left(\frac{LB_{adj}}{LB_{obs}}-1\right)}
\label{eq:evalueadj-flip}
\end{equation}

In the case where the limiting bound (\(LB_{obs}\)) is less than one, both \(LB_{obs}\) and \(LB_{adj}\) can be replaced with \(LB_{obs}^* = 1 / LB_{obs}\) and \(LB_{adj}^* = 1 / LB_{adj}\), respectively.

This Observed Covariate E-value quantifies the effect of moving the observed limiting bound (the confidence limit closest to the null of the effect observed \emph{without} a given confounder) to the adjusted limiting bound (the limiting bound of the effect \emph{with} the confounder). Practically, this can be calculated by using a metric we propose below, an \emph{observed bias effect}, for each measured confounder to quantify \(LB_{obs}\) along with the observed exposure-outcome effect adjusting for all observed confounders, \(LB_{adj}\). This adds context to the E-value, allowing it to be grounded in the observed covariates. This Observed Covariate E-value need not be limited to the effect of leaving out a single confounder. For example, it may be of interest to see the effect of leaving out a group of confounders. We will demonstrate this in the Example section below.

\hypertarget{observed-bias-effect}{%
\section{Observed bias effect}\label{observed-bias-effect}}

While calculating the impact of the observed covariates individually and in groups on the E-value scale is important for contextualizing the E-value sensitivity analysis, it is helpful to see the corresponding \emph{observed bias effects} on the original scale of the study result (Table 1).

The general idea is similar to the ``omitted variable bias'' discussed by Hosman et al.~(Hosman, Hansen, and Holland 2010) and ``calibrated sensitivity analysis'' as discussed by Hsu and Small (Hsu and Small 2013). Here, we are interested in how omitting each observed covariate, or a group of covariates, shifts the point estimate and 95\% confidence interval of the exposure-outcome effect. To calculate the observed bias effect, we first fit our model(s) as we normally would. In the case of an analysis that includes propensity score adjustment, for example, we fit the propensity score model and then the outcome model, estimating the exposure-outcome effect. We then repeat the entire process, leaving one covariate (or group of covariates) out at a time, and record the exposure effect and 95\% confidence interval each time. This triplet, the exposure effect for each updated process along with 95\% confidence interval, is the observed bias effect, demonstrating how the effect of interest would change had we not observed the covariate (or group of covariates) at hand. We then can use the limiting bound of these observed bias effects to calculate the Observed Covariate E-value.

Table 1:\label{tab:algo} Algorithm to calculate observed bias effect.

\begin{longtable}[]{@{}l@{}}
\toprule
\begin{minipage}[b]{0.77\columnwidth}\raggedright
\textbf{Algorithm to calculate \emph{observed bias effect}}\strut
\end{minipage}\tabularnewline
\midrule
\endhead
\begin{minipage}[t]{0.77\columnwidth}\raggedright
Let the full data be represented by \(\mathbf{D} = \{\mathbf{y}, \mathbf{z}, \mathbf{X}\}\),
where \(\mathbf{y}\) is the outcome, \(\mathbf{z}\) is the exposure of interest, and
\(\mathbf{X}\) is a matrix of the covariates, \(\mathbf{X} = \{\mathbf{x_1},...,\mathbf{x_k}\}\). \vspace{0.5cm}\strut
\end{minipage}\tabularnewline
\begin{minipage}[t]{0.77\columnwidth}\raggedright
Let \(\mathbf{G}\) be a matrix of matrices consisting of groups of covariates, \(\mathbf{g} = \{\mathbf{X}_g\}\), selected to demonstrate the bias associated with not including the group of covariates in the analysis, \(\mathbf{G} = \{\mathbf{g_1}, ..., \mathbf{g_l}\}\).
\vspace{0.5cm}\strut
\end{minipage}\tabularnewline
\begin{minipage}[t]{0.77\columnwidth}\raggedright
1. Perform the full analysis on \(\mathbf{D}\), for example fit the propensity score model as well as the outcome model, including all pre-specified covariates. Save the exposure-outcome effect and confidence interval, \((E_\mathbf{D}, LCL_\mathbf{D}, UCL_\mathbf{D})\).
\vspace{0.5cm}\strut
\end{minipage}\tabularnewline
\begin{minipage}[t]{0.77\columnwidth}\raggedright
2. For each \(i\) in \(\mathbf{X} = \{\mathbf{x_1},...,\mathbf{x_k}\}\), refit the the same analysis as (1) on \(\mathbf{D}_{-\mathbf{x}_i}\), leaving \(\mathbf{x}_i\) out of the data frame, \(\mathbf{D}\). Save the \emph{observed bias effect}, the updated exposure-outcome effect and confidence interval, \(\left(E_{\mathbf{D}_{-\mathbf{x}_i}}, LCL_{\mathbf{D}_{-\mathbf{x}_i}}, UCL_{\mathbf{D}_{-\mathbf{x}_i}}\right)\).
\vspace{0.5cm}\strut
\end{minipage}\tabularnewline
\begin{minipage}[t]{0.77\columnwidth}\raggedright
3. For each \(j\) in \(\mathbf{G} =\{\mathbf{g_1}, ..., \mathbf{g_l}\}\), refit the same analysis as (1) on \(\mathbf{D}_{-\mathbf{g}_j}\), leaving \(\mathbf{g}_j\) out of the data frame, \(\mathbf{D}\). Save the \emph{observed bias effect}, the updated exposure-outcome effect and confidence interval, \(\left(E_{\mathbf{D}_{-\mathbf{g}_j}}, LCL_{\mathbf{D}_{-\mathbf{g}_j}}, UCL_{\mathbf{D}_{-\mathbf{g}_j}}\right)\).\strut
\end{minipage}\tabularnewline
\bottomrule
\end{longtable}

A key benefit of these observed bias effects is that they take into account the total impact of an observed covariate or group of covariates, incorporating three key elements:

\begin{enumerate}
\def\labelenumi{\arabic{enumi}.}
\tightlist
\item
  How imbalanced the covariate is in the exposure
\item
  The association between the covariate and the outcome
\item
  How independent the covariate is from the remaining measured covariates
\end{enumerate}

For a confounder to be impactful, the interplay between these three elements is important. For example, a covariate that is highly imbalanced between exposure groups but has a very small association with the outcome is not likely to cause a large impact. Similarly, a covariate that is highly imbalanced, has a strong relationship with the outcome, but is highly dependent on the remaining measured covariates will also have a small impact.

\hypertarget{observed-bias-plot}{%
\section{Observed bias plot}\label{observed-bias-plot}}

In order to visualize these Observed Covariate E-values and observed bias effects, we propose an \emph{observed bias plot}. We plot the observed bias effect for each observed covariate alongside the associated Observed Covariate E-value, that is the E-value for moving the observed limiting bound (the limiting bound of the effect observed without the confounder) to the adjusted limiting bound (the limiting bound of the effect with the confounder), using Equation \eqref{eq:evalueadj}. This adds context to the E-value, allowing it to be grounded in the observed covariates. Essentially, it allows the researcher to re-conduct the analysis, leaving out a covariate or group of covariates. This left half of this plot demonstrates the effect this would have on the observed relationship between the exposure and outcome. The right side shows the Observed Covariate E-value, that is the E-value the study would have should the dropped covariate(s) be the only ``unmeasured confounder(s)'' that exist.

Similar to the Observed Covariate E-value and observed bias effect, these observed bias plots need not be limited to the effect of leaving out each confounder one at a time. For example, it may be of interest to see the effect of leaving out a group of confounders. In the example below, we leave out a number of groups including all measured covariates or all physiological measurements to demonstrate how that would have changed our result. In addition, we can add a shifted effect for a hypothetical unmeasured confounder that would tip this analysis, i.e.~bring the limiting bound of the effect to 1, as well as a hypothetical unmeasured confounder that would bring the point estimate to 1. The Observed Covariate E-values associated with these hypothetical shifted effects are exactly the E-values proposed by VanderWeele and Ding (VanderWeele and Ding 2017).

\hypertarget{example}{%
\section{Example}\label{example}}

To demonstrate the Observed Covariate E-value, observed bias effect, and the associated observed bias plot, we will use the Right Heart Catheterization (RHC) \href{http://biostat.mc.vanderbilt.edu/wiki/Main/DataSets}{dataset}, originally used in Connors et al (Connors et al. 1996). Connors et al.~provides an excellent example of a clinically thoughtful, quantified, contextualized sensitivity analysis. It included three key parts. First, Connors et al.~queried 13 content experts on what the ten most important factors were for choosing the exposure of interest, RHC, and included them in the propensity score along with other potential confounders. Second, a quantified sensitivity analysis to unmeasured confounding was performed using the methods put forth by Rosenbaum and Rubin (Rosenbaum and Rubin 1983). Third, the researchers conducted an additional sensitivity analysis, leaving out the covariates that were the four largest predictors in the propensity score. This additional analysis demonstrates how sensitive the study result is to covariates that are large predictors of the exposure of interest. Because the strongest predictors of either the exposure or outcome alone may not be the confounders that have the strongest impact on the study result, our method emphasizes looking at all of the covariates.

This study assessed the effectiveness of right heart catheterization (RHC) in the initial care of critically ill patients. The cohort contains 5,735 patients, 2,184 in the treatment group (RHC) and 3,551 in the control group (no RHC). This is a particularly interesting observational study, as it demonstrated a result counter to previously published recommendations for the use of RHC. The original analysis included 50 covariates used to estimate the propensity of being assigned to RHC. For demonstration purposes, we choose 20 to use here. We use demographics (age, sex), comorbidities (upper GI bleeding, renal disease, transfer status), physiological measurements taken on day 1 (bilirubin, hematocrit, white blood cell count, mean blood pressure, pH, PaO2/FiO2 ratio, albumin, respiratory rate, PaCO2, heart rate), diagnosis categories (Neurology and Hematology), APACHE score, the SUPPORT model estimate of the probability of surviving two months, and DNR status on day 1. Please see Connors et al.~for the fully adjusted analysis and clinical interpretation of the RHC effect (Connors et al. 1996). After fitting the propensity score model, we examine the balance using standardized mean differences via a Love plot (Love 2002; Hansen and Fredrickson 2014). Using the propensity scores, we construct overlap weights (Li, Morgan, and Zaslavsky 2018) for each individual and perform a weighted survival analysis estimating the effect of right heart catheterization on 30 day survival, adjusting for all 20 covariates. We then repeat these analysis steps, both fitting the propensity score model and the weighted outcome model, leaving out one covariate at a time. Additionally, we examine the impact of leaving out all covariates, APACHE score and the SUPPORT model estimate of the probability of surviving two months, all physiological measurements (bilirubin, hematocrit, white blood cell count, mean blood pressure, pH, PaO2/FiO2 ratio, albumin, respiratory rate, PaCO2, heart rate), or the combination of all physiological measurements, APACHE score and the SUPPORT probability of surviving two months. Each time we estimate the observed bias effect, the effect of the exposure, right heart catheterization, on the outcome, 30 day survival, and compare it to the estimate from the fully specified analysis.

\begin{figure}

{\centering \includegraphics{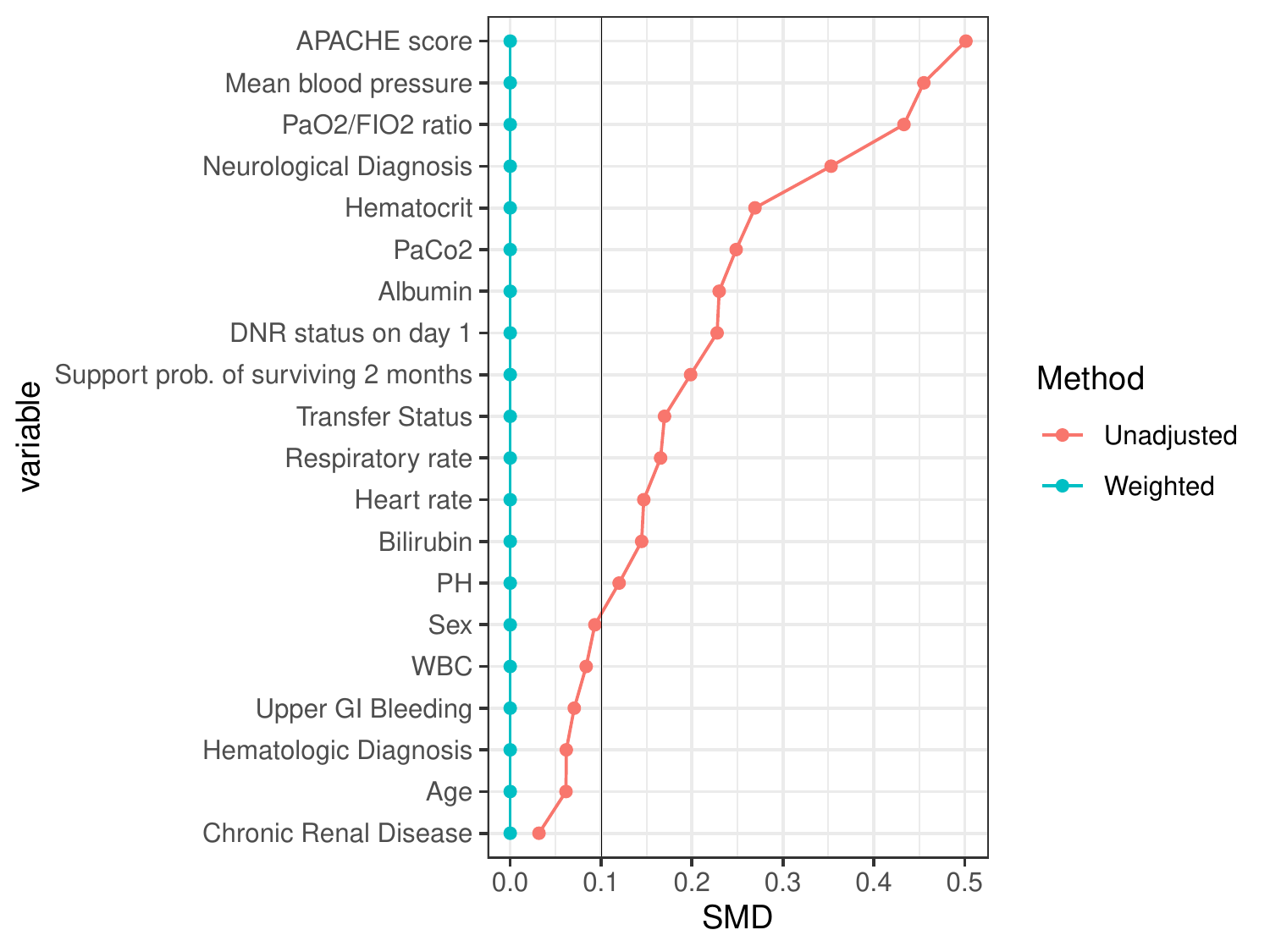} 

}

\caption[Love plot.]{Love plot. This displays the standardized mean difference between the exposed and unexposed groups before (red) and after (blue) propensity score weighting. The vertical line at 0.1 represents the "rule of thumb" for an acceptable standardized mean difference.}\label{fig:smd}
\end{figure}

\begin{figure}

{\centering \includegraphics{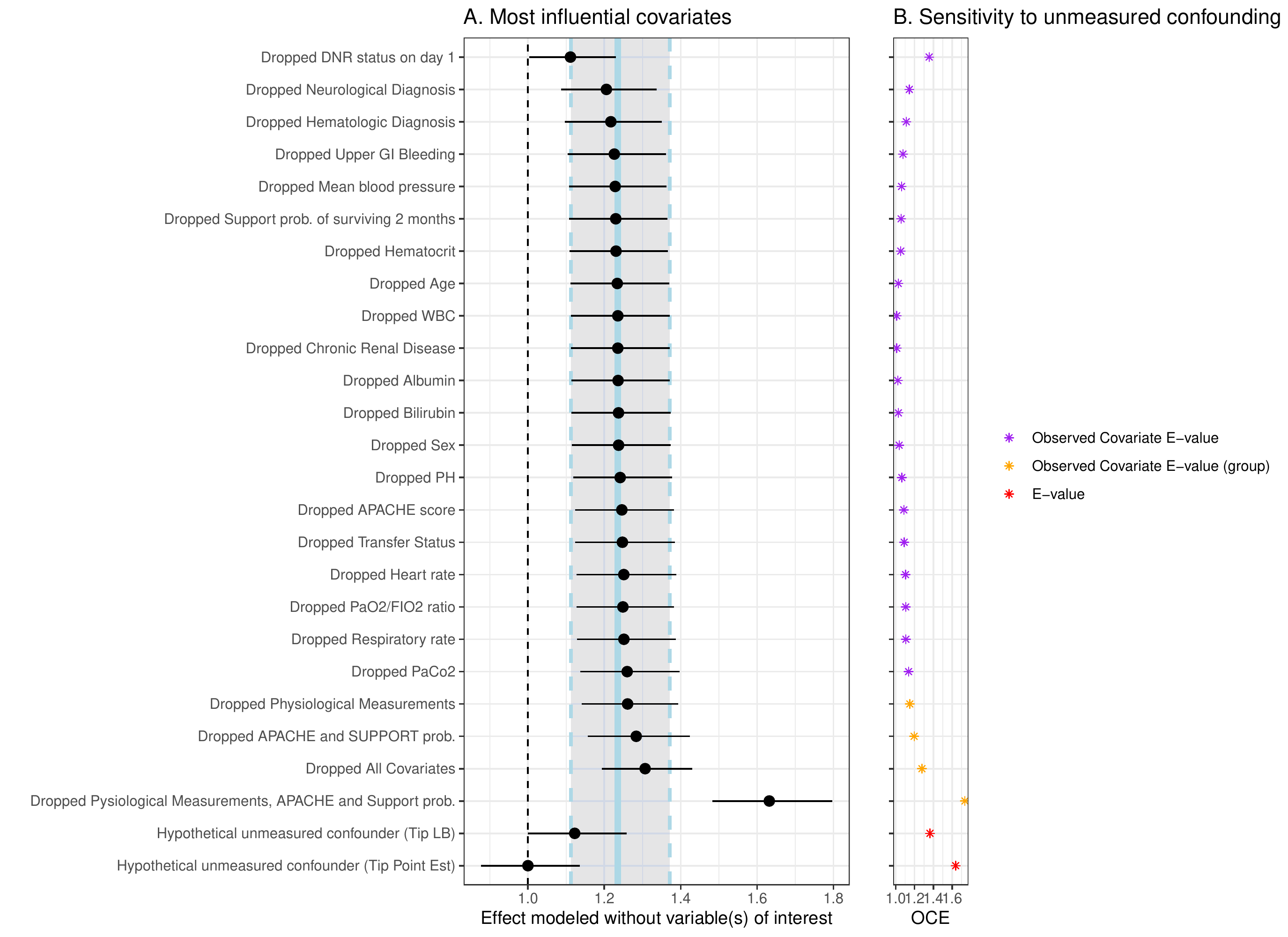} 

}

\caption[Observed bias plot]{Observed bias plot. Panel A displays the observed bias effects and Panel B displays the Observed Covariate E-values. This demonstrates the observed bias effects and Observed Covariate E-values for all covariates included in the analysis. Additionally, it examines the observed bias effects and Observed Covariate E-values for dropping all measured covariates, APACHE and Support probability, all physiological measurements, and all physiological measurements and APACHE and Support probability. Finally, we include what the observed bias effect of a hypothetical unmeasured confounder that would tip the analysis at the lower bound as well as a hypothetical unmeasured confounder that would tip the analysis at the point estimate, along with the corresponding E-values. This demonstrates how the hazard ratio and 95\% confidence interval of RHC on 30 day survival changes if each covariate, or group of covariates, were unobserved. The black dashed line at 1 represents the null. The solid blue line is the hazard ratio for RHC in the full model (1.24). The blue dashed lines filled with the light blue region represent the 95\% confidence interval for the association between RHC and 30 day survival in the full model (1.11, 1.37).}\label{fig:bias}
\end{figure}

Figure \ref{fig:smd} displays the Love plot and Figure \ref{fig:bias} displays the observed bias plot. The observed effect of RHC on 30 day survival is 1.24 (95\% CI: 1.11, 1.37). This is displayed as the blue line and shaded region in Figure \ref{fig:bias} as well as in Table \ref{tab:tab1}. Table \ref{tab:tab1} also displays each measured covariates association with the outcome, adjusting for all other covariates.

\begin{table}

\caption{\label{tab:tab1}The association with 30 day survival, adjusting for all other covariates.}
\centering
\begin{tabular}[t]{llcc}
\toprule
  & Hazard Ratio & 95\% LCL & 95\% UCL\\
\midrule
RHC & 1.24 & 1.11 & 1.37\\
Chronic Renal Disease & 1.05 & 0.80 & 1.37\\
Upper GI Bleeding & 1.58 & 1.23 & 2.03\\
Transfer Status & 1.30 & 1.11 & 1.52\\
APACHE score & 1.00 & 1.00 & 1.01\\
\addlinespace
WBC & 1.00 & 1.00 & 1.00\\
Heart rate & 1.00 & 1.00 & 1.00\\
PaO2/FIO2 ratio & 1.00 & 1.00 & 1.00\\
Albumin & 0.98 & 0.92 & 1.04\\
Hematocrit & 1.00 & 0.99 & 1.01\\
\addlinespace
Bilirubin & 1.03 & 1.02 & 1.04\\
Mean blood pressure & 1.00 & 1.00 & 1.00\\
PaCo2 & 0.99 & 0.99 & 1.00\\
DNR status on day 1 & 2.59 & 2.22 & 3.02\\
PH & 0.62 & 0.34 & 1.14\\
\addlinespace
Respiratory rate & 1.00 & 0.99 & 1.00\\
Neurological Diagnosis & 1.40 & 1.17 & 1.68\\
Hematologic Diagnosis & 1.39 & 1.16 & 1.67\\
Sex & 1.07 & 0.97 & 1.19\\
Age & 1.00 & 1.00 & 1.01\\
\addlinespace
Support prob. of surviving 2 months & 0.08 & 0.06 & 0.11\\
\bottomrule
\end{tabular}
\end{table}

\hypertarget{interpretation}{%
\subsection{Interpretation}\label{interpretation}}

Often the impact of observed covariates is examined separately between the exposure and outcome. Methods such as the standardized mean difference, visualized via the Love Plot in Figure \ref{fig:smd}, are used to demonstrate the relationship between the observed covariates and the exposure. Tables such as Table \ref{tab:tab1} are used to demonstrate the relationship between the observed covariates and the outcome. A benefit of the observed bias effects, as displayed in the observed bias plot (Figure \ref{fig:bias}), is that they quantify the total impact of the observed covariate, incorporating three key elements: 1. the relationship between the exposure and the observed covariate, 2. the relationship between the outcome and the observed covariate, and 3. the relationship between the observed covariate and the remaining measured covariates. If we were to only examine Figure \ref{fig:smd} alone, we would surmise that the observed covariate that has the largest impact is APACHE score, since it has the greatest imbalance in the exposure. If we were to only examine Table \ref{tab:tab1}, we would infer that the observed covariate that has the largest impact is the SUPPORT probability of surviving two months, since it has the strongest relationship with the outcome. However, when incorporating all three elements that determine the most influential covariate using the observed bias plot (Figure \ref{fig:bias}), we see that neither of these covariates have a very strong impact on the overall result of the model. In fact, DNR status on day 1 has a much larger impact. This is due to three things: 1. the imbalance between the exposure groups, 2. the magnitude of the relationship between DNR status on day 1 and the outcome, and 3. the relative independence of this variable from the other observed covariates. Both APACHE score and the SUPPORT probability of surviving two months have little impact when dropped from the analysis because the remaining covariates can capture the effect of those covariates. DNR status on day 1, however, is not well explained by the remaining covariates, and therefore is quite influential. This exploration can help a researcher better contextualize and conceive of a potential unmeasured confounder that could impact their analysis.

Given the observed lower bound of 1.11, the associated E-value is 1.36. Because our result is a hazard ratio, and our event rate is relatively common (33.44\%), we are using the transformation proposed by VanderWeele and Ding, as discussed in the E-value section above (VanderWeele and Ding 2017). Examining Figure \ref{fig:bias}, we can add some context to this value. Of the individual covariates, only DNR status has an Observed Covariate E-value that is close to VanderWeele and Ding's E-value (DNR status Observed Covariate E-value: 1.36). This implies that we would need to be missing an additional independent covariate akin to DNR status on day 1 in order to tip our analysis. In practice, we are treating our fully specified model as the ``truth'' and viewing the impact of having a single unmeasured confounder, in this case DNR status on day 1. If we had conducted our study missing DNR status on day 1, we would have underestimated the ``true'' effect, the effect including all measured covariates. We can then use this E-value to better understand the E-value for our overall analysis. A variable similar in size and magnitude to DNR status on day 1, but that would pull the result in the opposite direction (that is, pull the result \emph{away} from the null) would almost tip our analysis, since the Observed Covariate E-value for DNR status on day 1 is close to our study's E-value. Of note, the unmeasured confounder that would tip our analysis would be akin to DNR status on day 1, but in this framing, we \emph{would also} be adjusting for DNR status on day 1, since that was adjusted for in our full analysis, as the E-value describes the strength of association need to tip the analysis conditional on all measured covariates. A variable like DNR status on day 1 is not the only way to tip our analysis, the E-value itself describes that an unmeasured confounder with an association of 1.36 with both the exposure and outcome would also tip our analysis. Being able to draw the connection between an arbitrary unmeasured confounder and an actual measured confounder can potentially help the researcher assess it's plausibility. Similarly, we can look to the groups of covariates to help us understand the potential impact of several small unmeasured confounders. The observed bias effects (Panel A on Figure \ref{fig:bias}), allow us to see the direction of the bias introduced by dropping groups of covariates. Dropping all covariates, that is fitting the survival model using including only the exposure, results in a stronger effect. The models that drop APACHE score and the SUPPORT probability of surviving two months, all physiological measurements, and the combination of all physiological measurements, APACHE score, and the SUPPORT probability of surviving two months also result in stronger effects. We can then look at Panel B on Figure \ref{fig:bias} to see the Observed Covariate E-values for dropping these groups of covariates. These help the researcher ponder: what if there were a group of unmeasured confounders of the same magnitude as those observed here, but that pulled the result towards the null - would they be strong enough to tip the result? In the case of the first three groups - the group of all covariates, APACHE score and the SUPPORT probability of surviving two months, and the group of physiological measurements, the answer is \emph{no}. If all physiological measurements, APACHE score, and the SUPPORT probability of surviving two months were all dropped from the model, however, the impact is much greater. If a set of unmeasured confounders such as these were of the same magnitude, they would be sufficient to tip this analysis. This particular case is a nice illustration of the interplay with these covariates. Dropping \emph{all} covariates from the model does not seem to change the result too substantially, however dropping some (in this case 12 of the 20, 10 physiological measurements, APACHE score, and the SUPPORT probability of surviving two months) has a large impact on the result. This suggests both that the combined impact of these covariates is substantial. Dropping the APACHE score and the SUPPORT probability alone has little impact. Similarly dropping all physiological measurements has little impact. And yet, dropping the combination of APACHE score, the SUPPORT probability, and all physiological measurements has a substantial impact on the observed effect. This suggests that perhaps SUPPORT probability and APACHE score account for a similar relationship as the 10 physiological measurements, so as long as one of the two groups is included, the impact on the overall estimates is minimal. Removing all 12 covariates, however, results in the large impact that is seen. The nuance that this plot provides can help the researcher carefully consider the analysis at hand as well as the potential for how things could be misrepresented should there be some lingering unmeasured confounding.

All calculations and plots were created in R using the \textbf{tipr} package. For full code, see Supplemental Appendix A.

\hypertarget{discussion}{%
\section{Discussion}\label{discussion}}

The current rarity of quantified sensitivity to unmeasured confounding analyses, even in top tier medical journals, emphasizes the need for methods that can gain traction with medical researchers. We believe the E-value put forth by VanderWeele and Ding has the potential to achieve this with its ease of calculation. The limitations in interpreting the E-value can be remedied by grounding it in the effects of the observed covariates, via the Observed Covariate E-value, observed bias effect, and observed bias plot. Taken as a whole, this contextualized, quantified sensitivity analysis focuses readers on the quality of the covariate adjustment the study was able to perform and on making a reasoned judgment on the likelihood of sufficiently strong confounding remaining unaccounted for.

In addition to demonstrating the utility of our new method, we seek to portray best practices for conducting observational study research, such as including a Love Plot to examine pre-treatment covariate balance pre- and post-propensity score adjustment (Austin and Stuart 2015; Joffe et al. 2004). Not only does this plot exhibit current best practices, it also demonstrate that sensitivity analyses including observed covariates ought to consist of more than just covariates that are imbalanced in the exposure. We illustrate that the covariates that are the most imbalanced, such as APACHE score, mean blood pressure, PaO2/FIO2 ratio, and neurological diagnosis, do not necessarily have the largest impact on the overall analysis, as seen in the observed bias plot. The sensitivity to unmeasured confounding analysis proposed by this paper and encapsulated in the observed bias plot offers a succinct way to summarize the impact of all covariates and selected groups of covariates, while providing contextual comparisons for the E-value sensitivity analysis.

\hypertarget{references}{%
\section*{References}\label{references}}
\addcontentsline{toc}{section}{References}

\hypertarget{refs}{}
\begin{cslreferences}
\leavevmode\hypertarget{ref-Austin:2015co}{}%
Austin, Peter C, and Elizabeth A Stuart. 2015. ``Moving towards best practice when using inverse probability of treatment weighting (IPTW) using the propensity score to estimate causal treatment effects in observational studies.'' \emph{Statistics in Medicine} 34 (28): 3661--79.

\leavevmode\hypertarget{ref-Bross}{}%
Bross, Irwin D J. 1966. ``Spurious effects from an extraneous variable.'' \emph{Journal of Chronic Diseases} 19 (6): 637--47.

\leavevmode\hypertarget{ref-Brumback:2004gk}{}%
Brumback, Babette A, Miguel A Hernán, Sebastien J P A Haneuse, and James M Robins. 2004. ``Sensitivity analyses for unmeasured confounding assuming a marginal structural model for repeated measures.'' \emph{Statistics in Medicine} 23 (5): 749--67.

\leavevmode\hypertarget{ref-Connors:1996un}{}%
Connors, Alfred F, Theodore Speroff, Neal V Dawson, Charles Thomas, Frank E Harrell, Douglas Wagner, Norman Desbiens, et al. 1996. ``The Effectiveness of Right Heart Catheterization in the Initial Care of Critically Ill Patients.'' \emph{Jama} 276 (11): 889--97.

\leavevmode\hypertarget{ref-Cornfield}{}%
Cornfield, J, W Haenszel, E C Hammond, A M Lilienfeld, M B Shimkin, and E L Wynder. 1959. ``Smoking and lung cancer: recent evidence and a discussion of some questions.'' \emph{Journal of the National Cancer Institute} 22 (1): 173--203.

\leavevmode\hypertarget{ref-Ding}{}%
Ding, Peng, and Tyler J VanderWeele. 2016. ``Sensitivity Analysis Without Assumptions.'' \emph{Epidemiology} 27 (3): 368--77.

\leavevmode\hypertarget{ref-Greenland:1998}{}%
Greenland, Sander. 1998. ``The sensitivity of a sensitivity analysis.'' \emph{1997 Proceedings of the Biometrics Section American Statistical Association}, 19--21.

\leavevmode\hypertarget{ref-Greenland:2001kf}{}%
---------. 2001. ``Sensitivity Analysis, Monte Carlo Risk Analysis, and Bayesian Uncertainty Assessment.'' \emph{Risk Analysis} 21 (4): 579--84.

\leavevmode\hypertarget{ref-Greenland:2003ha}{}%
---------. 2003. ``The Impact of Prior Distributions for Uncontrolled Confounding and Response Bias.'' \emph{Journal of the American Statistical Association} 98 (461): 47--54.

\leavevmode\hypertarget{ref-Greenland:2005gb}{}%
---------. 2005. ``Multiple-bias modelling for analysis of observational data.'' \emph{Journal of the Royal Statistical Society: Series A (Statistics in Society)} 168 (2): 267--306.

\leavevmode\hypertarget{ref-Groenwold:2008hj}{}%
Groenwold, Rolf H H, Anna M M Van Deursen, Arno W Hoes, and Eelko Hak. 2008. ``Poor Quality of Reporting Confounding Bias in Observational Intervention Studies: A Systematic Review.'' \emph{Annals of Epidemiology} 18 (10): 746--51.

\leavevmode\hypertarget{ref-Hansen:2014wn}{}%
Hansen, Ben B, and Mark M Fredrickson. 2014. ``Omitted Variable Sensitivity Analysis with the Annotated Love Plot.'' \emph{Society for Research on Educational Effectiveness}.

\leavevmode\hypertarget{ref-Hosman:2010fr}{}%
Hosman, C A, B B Hansen, and P W Holland. 2010. ``The Sensitivity of Linear Regression Coefficients' Confidence Limits to the Omission of a Confounder.'' \emph{The Annals of Applied Statistics} 4 (2): 849--70.

\leavevmode\hypertarget{ref-Hsu:2013ky}{}%
Hsu, Jesse Y, and Dylan S Small. 2013. ``Calibrating Sensitivity Analyses to Observed Covariates in Observational Studies.'' \emph{Biometrics} 69 (4): 803--11.

\leavevmode\hypertarget{ref-Joffe:2004ho}{}%
Joffe, Marshall M, Thomas R Ten Have, Harold I Feldman, and Stephen E Kimmel. 2004. ``Model Selection, Confounder Control, and Marginal Structural Models.'' \emph{The American Statistician} 58 (4): 272--79.

\leavevmode\hypertarget{ref-lash2011applying}{}%
Lash, Timothy L, Matthew P Fox, and Aliza K Fink. 2011. \emph{Applying Quantitative Bias Analysis to Epidemiologic Data}. Springer Science \& Business Media.

\leavevmode\hypertarget{ref-10.1093ux2fijeux2fdyu149}{}%
Lash, Timothy L, Matthew P Fox, Richard F MacLehose, George Maldonado, Lawrence C McCandless, and Sander Greenland. 2014. ``Good practices for quantitative bias analysis.'' \emph{International Journal of Epidemiology} 43 (6): 1969--85. \url{https://doi.org/10.1093/ije/dyu149}.

\leavevmode\hypertarget{ref-Li:2016jk}{}%
Li, Fan, Kari Lock Morgan, and Alan M Zaslavsky. 2018. ``Balancing Covariates via Propensity Score Weighting.'' \emph{Journal of the American Statistical Association} 113 (521): 390--400.

\leavevmode\hypertarget{ref-Lin}{}%
Lin, D Y, B M Psaty, and R A Kronmal. 1998. ``Assessing the sensitivity of regression results to unmeasured confounders in observational studies.'' \emph{Biometrics} 54 (3): 948--63.

\leavevmode\hypertarget{ref-Love:2002vf}{}%
Love, Thomas E. 2002. ``Displaying Covariate Balance After Adjustment for Selection Bias.'' In \emph{Section on Health Policy Statistics, Joint Statistical Meetings, New York}. Vol. 11.

\leavevmode\hypertarget{ref-McCandless:2007he}{}%
McCandless, Lawrence C, Paul Gustafson, and Adrian Levy. 2007. ``Bayesian sensitivity analysis for unmeasured confounding in observational studies.'' \emph{Statistics in Medicine} 26 (11): 2331--47.

\leavevmode\hypertarget{ref-McCandless:2008fd}{}%
McCandless, Lawrence C, Paul Gustafson, and Adrian R Levy. 2008. ``A sensitivity analysis using information about measured confounders yielded improved uncertainty assessments for unmeasured confounding.'' \emph{Journal of Clinical Epidemiology} 61 (3): 247--55.

\leavevmode\hypertarget{ref-Robins:2000es}{}%
Robins, James M, Andrea Rotnitzky, and Daniel O Scharfstein. 2000. ``Sensitivity Analysis for Selection bias and unmeasured Confounding in missing Data and Causal inference models.'' In \emph{Statistical Models in Epidemiology, the Environment, and Clinical Trials}, 1--94. New York, NY: Springer New York.

\leavevmode\hypertarget{ref-Rosenbaum:1983}{}%
Rosenbaum, P. R., and D. B. Rubin. 1983. ``Assessing Sensitivity to an Unobserved Binary Covariate in an Observational Study with Binary Outcome.'' \emph{Journal of the Royal Statistical Society. Series B (Methodological)} 45 (2): 212--18.

\leavevmode\hypertarget{ref-Schlesselman}{}%
Schlesselman, J J. 1978. ``Assessing effects of confounding variables.'' \emph{American Journal of Epidemiology} 108 (1): 3--8.

\leavevmode\hypertarget{ref-Schneeweiss:2006bc}{}%
Schneeweiss, Sebastian. 2006. ``Sensitivity analysis and external adjustment for unmeasured confounders in epidemiologic database studies of therapeutics.'' \emph{Pharmacoepidemiology and Drug Safety} 15 (5): 291--303.

\leavevmode\hypertarget{ref-vanBelle}{}%
Van Belle, Gerald. 2011. \emph{Statistical Rules of Thumb}. Vol. 699. John Wiley \& Sons.

\leavevmode\hypertarget{ref-VanderWeele:2008uq}{}%
VanderWeele, Tyler J. 2008a. ``Sensitivity Analysis: Distributional Assumptions and Confounding Assumptions.'' \emph{Biometrics} 64 (2): 645--49.

\leavevmode\hypertarget{ref-VanderWeele:2008dz}{}%
---------. 2008b. ``The Sign of the Bias of Unmeasured Confounding.'' \emph{Biometrics} 64 (3): 702--6.

\leavevmode\hypertarget{ref-VanderWeele:2013ho}{}%
---------. 2013. ``Unmeasured confounding and hazard scales: sensitivity analysis for total, direct, and indirect effects.'' \emph{European Journal of Epidemiology} 28 (2): 113--17.

\leavevmode\hypertarget{ref-VanderWeele:2011es}{}%
VanderWeele, Tyler J, and Onyebuchi A Arah. 2011. ``Bias Formulas for Sensitivity Analysis of Unmeasured Confounding for General Outcomes, Treatments, and Confounders.'' \emph{Epidemiology} 22 (1): 42--52.

\leavevmode\hypertarget{ref-VanderWeele:2017ki}{}%
VanderWeele, Tyler J, and Peng Ding. 2017. ``Sensitivity Analysis in Observational Research: Introducing the E-Value.'' \emph{Annals of Internal Medicine} 167 (4): 268--74.

\leavevmode\hypertarget{ref-VanderWeele:2008uc}{}%
VanderWeele, Tyler J, Miguel A Hernán, and James M Robins. 2008. ``Causal directed acyclic graphs and the direction of unmeasured confounding bias.'' \emph{Epidemiology} 19 (5): 720--28.

\leavevmode\hypertarget{ref-VanderWeele:2012bd}{}%
VanderWeele, Tyler J, Bhramar Mukherjee, and Jinbo Chen. 2012. ``Sensitivity analysis for interactions under unmeasured confounding.'' \emph{Statistics in Medicine} 31 (22): 2552--64.
\end{cslreferences}

\end{document}


\maketitle

\newthought{Here we are going to demonstrate} how to calculate the
Observed Covariate E-Value and create the accompanying observed bias
plot. We are going to use the same example highlighted in the main text,
using the Right Heart Catheterization (RHC)
\href{http://biostat.mc.vanderbilt.edu/wiki/Main/DataSets}{dataset},
originally used in \citet{Connors:1996un}. The first section, \emph{Data
processing / fitting the full model} simply demonstrates how to conduct
one type of analysis on these data. The second section
\emph{\textbf{tipr} Demonstration} shows how to use our package,
\textbf{tipr}, to calculate Observed Covariate E-values as well as an
observed bias plot using the analysis set up in the first section.

\hypertarget{data-processing-fitting-the-full-model}{%
\section{Data processing / fitting the full
model}\label{data-processing-fitting-the-full-model}}

We need four packages to run this analysis. The \textbf{tipr} packages
can be installed from GitHub using the following code.

\begin{Shaded}
\begin{Highlighting}[]
\CommentTok{\# install.packages("remotes")}
\NormalTok{remotes}\OperatorTok{::}\KeywordTok{install\_github}\NormalTok{(}\StringTok{"LucyMcGowan/tipr"}\NormalTok{,}
                        \DataTypeTok{ref =} \StringTok{"observed{-}bias{-}plot"}\NormalTok{)}
\end{Highlighting}
\end{Shaded}

\begin{marginfigure}
\hypertarget{packages}{%
\subsection{Packages}\label{packages}}

\begin{itemize}
\tightlist
\item
  \textbf{tipr}: to calculate the Observed Covariate E-value and set up
  the data for the observed bias plot.
\item
  \textbf{tidyverse}: to process the data. \emph{Note: This could also
  be accomplished without this package, we are simply demonstrating one
  way to to this.}
\item
  \textbf{survey}: to implement propensity score weighting.
\item
  \textbf{survival}: to complete the outcome analysis.
\item
  \textbf{gridExtra}: to combine plots.
\end{itemize}
\end{marginfigure}

\begin{Shaded}
\begin{Highlighting}[]
\KeywordTok{library}\NormalTok{(tipr)}
\KeywordTok{library}\NormalTok{(tidyverse)}
\KeywordTok{library}\NormalTok{(survey)}
\KeywordTok{library}\NormalTok{(survival)}
\KeywordTok{library}\NormalTok{(gridExtra)}
\end{Highlighting}
\end{Shaded}

The RHC data can be obtained from the Vanderbilt Datasets Wiki page.

\begin{Shaded}
\begin{Highlighting}[]
\NormalTok{rhc \textless{}{-}}\StringTok{ }\KeywordTok{read\_csv}\NormalTok{(}\StringTok{"http://biostat.mc.vanderbilt.edu/wiki/pub/Main/DataSets/rhc.csv"}\NormalTok{)}
\end{Highlighting}
\end{Shaded}

\hypertarget{data-cleaning}{%
\subsection{Data Cleaning}\label{data-cleaning}}

Here our exposure is \texttt{swang1}, our 30 day survival time is
\texttt{t3d30} and our event variable is \texttt{dth30}. We are going to
update my exposure variable to take the values \texttt{0} and \texttt{1}
rather than \texttt{RHC} and \texttt{No\ RHC} and rename this variable
\texttt{exposure}. We will will also rename \texttt{t3d30} \texttt{time}
and update \texttt{dth30} to take values \texttt{0} and \texttt{1},
calling this new variable \texttt{event}.

\begin{Shaded}
\begin{Highlighting}[]
\NormalTok{rhc \textless{}{-}}\StringTok{ }\NormalTok{rhc }\OperatorTok{\%\textgreater{}\%}
\StringTok{  }\KeywordTok{mutate}\NormalTok{(}
    \DataTypeTok{exposure =} \KeywordTok{case\_when}\NormalTok{(}
\NormalTok{      swang1 }\OperatorTok{==}\StringTok{ "RHC"} \OperatorTok{\textasciitilde{}}\StringTok{ }\DecValTok{1}\NormalTok{,}
      \OtherTok{TRUE} \OperatorTok{\textasciitilde{}}\StringTok{ }\DecValTok{0}
\NormalTok{    ),}
    \DataTypeTok{time =}\NormalTok{ t3d30,}
    \DataTypeTok{event =} \KeywordTok{case\_when}\NormalTok{(}
\NormalTok{      dth30 }\OperatorTok{==}\StringTok{ "Yes"} \OperatorTok{\textasciitilde{}}\StringTok{ }\DecValTok{1}\NormalTok{,}
      \OtherTok{TRUE} \OperatorTok{\textasciitilde{}}\StringTok{ }\DecValTok{0}
\NormalTok{    )}
\NormalTok{  )}
\end{Highlighting}
\end{Shaded}

We have chosen the following 20 variables to include in this analysis.

\begin{Shaded}
\begin{Highlighting}[]
\NormalTok{vars \textless{}{-}}\StringTok{ }\KeywordTok{c}\NormalTok{(}\StringTok{"renalhx"}\NormalTok{, }\StringTok{"gibledhx"}\NormalTok{, }\StringTok{"transhx"}\NormalTok{, }\StringTok{"aps1"}\NormalTok{, }\StringTok{"wblc1"}\NormalTok{,}
          \StringTok{"hrt1"}\NormalTok{, }\StringTok{"pafi1"}\NormalTok{, }\StringTok{"alb1"}\NormalTok{, }\StringTok{"hema1"}\NormalTok{, }\StringTok{"bili1"}\NormalTok{,}
          \StringTok{"meanbp1"}\NormalTok{, }\StringTok{"paco21"}\NormalTok{, }\StringTok{"dnr1"}\NormalTok{, }\StringTok{"ph1"}\NormalTok{, }\StringTok{"resp1"}\NormalTok{,}
          \StringTok{"neuro"}\NormalTok{, }\StringTok{"hema"}\NormalTok{, }\StringTok{"sex"}\NormalTok{, }\StringTok{"age"}\NormalTok{, }\StringTok{"surv2md1"}\NormalTok{)}
\end{Highlighting}
\end{Shaded}

\hypertarget{propensity-score-model}{%
\subsection{Propensity score model}\label{propensity-score-model}}

We fit the full propensity score model using a logistic regression with
all 20 variables included. We then add the propensity score \texttt{p}
to the \texttt{rhc} data frame as well as the \texttt{weight}
(\citet{Li:2016jk}). For this analysis, we are using a weighted outcome
model to incorporate the propensity score for demonstration purposes.
Other propensity score methods could be implemented (such as matching,
covariate adjustment, etc.) with the methods put forth in this paper.

\begin{Shaded}
\begin{Highlighting}[]
\NormalTok{ps\_frm \textless{}{-}}\StringTok{ }\KeywordTok{as.formula}\NormalTok{(}
  \KeywordTok{paste}\NormalTok{(}\StringTok{"exposure \textasciitilde{} "}\NormalTok{, }\KeywordTok{paste}\NormalTok{(vars, }\DataTypeTok{collapse =} \StringTok{"+"}\NormalTok{))}
\NormalTok{)}

\NormalTok{ps\_mod \textless{}{-}}\StringTok{ }\KeywordTok{glm}\NormalTok{(ps\_frm,}
              \DataTypeTok{family =} \KeywordTok{binomial}\NormalTok{(}\DataTypeTok{link =} \StringTok{"logit"}\NormalTok{),}
              \DataTypeTok{data =}\NormalTok{ rhc)}
\end{Highlighting}
\end{Shaded}

\begin{marginfigure}
\vspace{3cm}

\hypertarget{note}{%
\subsection{Note}\label{note}}

For this analysis, we use overlap weights (Li et al.~{[}2016{]}). The
methods put forth here could work with any appropriate weighting scheme.
\end{marginfigure}

\begin{Shaded}
\begin{Highlighting}[]
\NormalTok{rhc \textless{}{-}}\StringTok{ }\NormalTok{rhc }\OperatorTok{\%\textgreater{}\%}
\StringTok{  }\KeywordTok{mutate}\NormalTok{(}
    \DataTypeTok{p =} \KeywordTok{predict}\NormalTok{(ps\_mod, }\DataTypeTok{type =} \StringTok{"response"}\NormalTok{),}
    \DataTypeTok{weight =} \KeywordTok{case\_when}\NormalTok{(}
\NormalTok{      exposure }\OperatorTok{==}\StringTok{ }\DecValTok{1} \OperatorTok{\textasciitilde{}}\StringTok{ }\DecValTok{1} \OperatorTok{{-}}\StringTok{ }\NormalTok{p,}
      \OtherTok{TRUE} \OperatorTok{\textasciitilde{}}\StringTok{ }\NormalTok{p}
\NormalTok{    )}
\NormalTok{  )}
\end{Highlighting}
\end{Shaded}

Using the \textbf{survey} package, we create a survey design object that
incorporates the propensity score weights we have just created. We will
use this in the outcome model.

\begin{Shaded}
\begin{Highlighting}[]
\NormalTok{weight\_des \textless{}{-}}\StringTok{ }\KeywordTok{svydesign}\NormalTok{(}\DataTypeTok{ids =} \OperatorTok{\textasciitilde{}}\StringTok{ }\DecValTok{1}\NormalTok{,}
                        \DataTypeTok{data =}\NormalTok{ rhc,}
                        \DataTypeTok{weights =} \OperatorTok{\textasciitilde{}}\StringTok{ }\NormalTok{weight)}
\end{Highlighting}
\end{Shaded}

\hypertarget{outcome-model}{%
\subsection{Outcome model}\label{outcome-model}}

We fit the full outcome model with all 20 covariates included. We are
fitting a weighted survival model, using the \texttt{svycoxph()}
function from the \textbf{survey} package.

\begin{marginfigure}
\hypertarget{note}{%
\subsection{Note}\label{note}}

Since we are going to be re-running this outcome model multiple times
with different formulas and propensity score weights, we can create a
small function that will run this code each time. We will use this
later:

\begin{verbatim}
run_outcome_mod <- 
  function(formula, wt) {
    des <- svydesign(ids = ~ 1,
                     data = rhc,
                     weights = wt)
    svycoxph(formula = formula,
             design = des)
  }
\end{verbatim}
\end{marginfigure}

\begin{Shaded}
\begin{Highlighting}[]
\NormalTok{outcome\_frm \textless{}{-}}\StringTok{ }\KeywordTok{as.formula}\NormalTok{(}
  \KeywordTok{paste}\NormalTok{(}\StringTok{"Surv(time, event) \textasciitilde{} exposure + "}\NormalTok{,}
        \KeywordTok{paste}\NormalTok{(vars, }\DataTypeTok{collapse =} \StringTok{"+"}\NormalTok{))}
\NormalTok{)}

\NormalTok{outcome\_mod \textless{}{-}}\StringTok{ }\KeywordTok{svycoxph}\NormalTok{(}\DataTypeTok{formula =}\NormalTok{ outcome\_frm,}
                        \DataTypeTok{design =}\NormalTok{ weight\_des)}
\end{Highlighting}
\end{Shaded}

Now that we have completed the analysis, let's create a small data frame
of the results to use later.

\begin{Shaded}
\begin{Highlighting}[]
\NormalTok{full\_model \textless{}{-}}\StringTok{ }\KeywordTok{data.frame}\NormalTok{(}
  \DataTypeTok{point\_estimate =} \KeywordTok{exp}\NormalTok{(}\KeywordTok{coef}\NormalTok{(outcome\_mod)[}\StringTok{"exposure"}\NormalTok{]),}
  \DataTypeTok{lb =} \KeywordTok{exp}\NormalTok{(}\KeywordTok{confint}\NormalTok{(outcome\_mod, }\StringTok{"exposure"}\NormalTok{)[, }\DecValTok{1}\NormalTok{]),}
  \DataTypeTok{ub =} \KeywordTok{exp}\NormalTok{(}\KeywordTok{confint}\NormalTok{(outcome\_mod, }\StringTok{"exposure"}\NormalTok{)[, }\DecValTok{2}\NormalTok{])}
\NormalTok{)}
\NormalTok{full\_model}
\end{Highlighting}
\end{Shaded}

\begin{verbatim}
##          point_estimate      lb       ub
## exposure       1.235202 1.11277 1.371105
\end{verbatim}

The result from our full model is a HR for the exposure of 1.24 (1.11,
1.37).

\hypertarget{tipr-demonstration}{%
\section{\texorpdfstring{\textbf{tipr}
Demonstration}{tipr Demonstration}}\label{tipr-demonstration}}

Now that we have set up a basic demonstration of an analysis of the RHC
data, we will demonstrate how to calculate the Observed Covariate
E-values as well as create an observed bias plot.

As mentioned in the main text, one advantage of these methods is they
can be performed to examine how removing a \emph{single} observed
covariate would change the final result, as well as how removing a
\emph{group} of observed covariates would change the result. We are
going examine how removing each variable, one at a time, will affect the
final result, as well as how removing all measured covariates, APACHE
and Support probability, all physiological measurements, or all
physiological measurements and APACHE and Support probability affects
the final result.

\hypertarget{step-1.-choose-variable-groups-to-drop-simultaneously}{%
\subsection{Step 1. Choose variable groups to drop
simultaneously}\label{step-1.-choose-variable-groups-to-drop-simultaneously}}

We will first create a named list of the two groups of variables.

\begin{Shaded}
\begin{Highlighting}[]
\CommentTok{\# name our full vector of variables}
\KeywordTok{names}\NormalTok{(vars) \textless{}{-}}\StringTok{ }\NormalTok{vars}

\CommentTok{\# create a named list of groups}
\NormalTok{groups \textless{}{-}}\StringTok{ }\KeywordTok{list}\NormalTok{( }
  \StringTok{"All Covariates"}\NormalTok{ =}\StringTok{ }\KeywordTok{unname}\NormalTok{(vars),}
  \StringTok{"APACHE and Support prob."}\NormalTok{ =}\StringTok{ }\KeywordTok{c}\NormalTok{(}\StringTok{"aps1"}\NormalTok{, }\StringTok{"surv2md1"}\NormalTok{),}
  \StringTok{"All Physiological Measurements"}\NormalTok{ =}
\StringTok{    }\KeywordTok{c}\NormalTok{(}\StringTok{"hrt1"}\NormalTok{, }\StringTok{"wblc1"}\NormalTok{, }\StringTok{"pafi1"}\NormalTok{, }\StringTok{"alb1"}\NormalTok{, }\StringTok{"hema1"}\NormalTok{, }\StringTok{"bili1"}\NormalTok{, }
      \StringTok{"paco21"}\NormalTok{, }\StringTok{"meanbp1"}\NormalTok{, }\StringTok{"resp1"}\NormalTok{, }\StringTok{"ph1"}\NormalTok{),}
  \StringTok{"Physiological Measurements, APACHE, and Support prob."}\NormalTok{ =}\StringTok{ }
\StringTok{    }\KeywordTok{c}\NormalTok{(}\StringTok{"hrt1"}\NormalTok{, }\StringTok{"wblc1"}\NormalTok{, }\StringTok{"pafi1"}\NormalTok{, }\StringTok{"alb1"}\NormalTok{, }\StringTok{"hema1"}\NormalTok{, }\StringTok{"bili1"}\NormalTok{,}
      \StringTok{"paco21"}\NormalTok{, }\StringTok{"meanbp1"}\NormalTok{, }\StringTok{"resp1"}\NormalTok{, }\StringTok{"ph1"}\NormalTok{, }\StringTok{"aps1"}\NormalTok{, }\StringTok{"surv2md1"}\NormalTok{)}
\NormalTok{)}

\CommentTok{\# append into a single named list}
\NormalTok{drop\_list \textless{}{-}}\StringTok{ }\KeywordTok{append}\NormalTok{(vars, groups)}
\end{Highlighting}
\end{Shaded}

\hypertarget{step-2.-create-an-observed-bias-data-frame}{%
\subsection{Step 2. Create an observed bias data
frame}\label{step-2.-create-an-observed-bias-data-frame}}

Using the \texttt{observed\_bias\_tbl()} function from \textbf{tipr}, we
will create a data frame with six columns:

\begin{itemize}
\tightlist
\item
  \texttt{dropped}: The variable or group of variables that was dropped
  from the analysis
\item
  \texttt{type}: Explanation of \texttt{dropped}, whether it refers to a
  single covariate (\texttt{covariate}) or a group of covariates
  (\texttt{group})
\item
  \texttt{ps\_formula}: An updated propensity score formula, based on
  the dropped variables
\item
  \texttt{outcome\_formula}: An updated outcome formula, based on the
  dropped variables
\item
  \texttt{ps\_model}: A model object for the updated propensity score,
  refit using the full model provided
\item
  \texttt{p}: The updated propensity score
\end{itemize}

\begin{marginfigure}
\hypertarget{tipr-function}{%
\subsection{\texorpdfstring{\textbf{tipr}
function}{tipr function}}\label{tipr-function}}

The \texttt{observed\_bias\_tbl()} function takes three parameters:

\begin{itemize}
\tightlist
\item
  \texttt{ps\_mod}: The full propensity score model we fit above
\item
  \texttt{outcome\_mod}: The full outcome model we fit above
\item
  \texttt{drop\_list}: A named list of covariates or groups of
  covariates we would like to simultaneously drop from the analysis. If
  \texttt{NULL} (default), will default to dropping each covariate
  included in the model one at a time.
\end{itemize}
\end{marginfigure}

\begin{Shaded}
\begin{Highlighting}[]
\NormalTok{(o \textless{}{-}}\StringTok{ }\KeywordTok{observed\_bias\_tbl}\NormalTok{(ps\_mod, outcome\_mod, drop\_list))}
\end{Highlighting}
\end{Shaded}

\begin{verbatim}
## # A tibble: 24 x 6
##    dropped  type      ps_formula outcome_formula ps_model p            
##    <chr>    <chr>     <list>     <list>          <list>   <list>       
##  1 renalhx  covariate <formula>  <formula>       <glm>    <dbl [5,735]>
##  2 gibledhx covariate <formula>  <formula>       <glm>    <dbl [5,735]>
##  3 transhx  covariate <formula>  <formula>       <glm>    <dbl [5,735]>
##  4 aps1     covariate <formula>  <formula>       <glm>    <dbl [5,735]>
##  5 wblc1    covariate <formula>  <formula>       <glm>    <dbl [5,735]>
##  6 hrt1     covariate <formula>  <formula>       <glm>    <dbl [5,735]>
##  7 pafi1    covariate <formula>  <formula>       <glm>    <dbl [5,735]>
##  8 alb1     covariate <formula>  <formula>       <glm>    <dbl [5,735]>
##  9 hema1    covariate <formula>  <formula>       <glm>    <dbl [5,735]>
## 10 bili1    covariate <formula>  <formula>       <glm>    <dbl [5,735]>
## # ... with 14 more rows
\end{verbatim}

\begin{marginfigure}
\hypertarget{note}{%
\subsection{Note}\label{note}}

Notice some of these columns are \texttt{list} columns. This allows us
to include full model objects, formulas, or all of the propensity scores
in a single row. In order to deal with these list columns, we will often
use the \texttt{map} functions from the \textbf{purrr} package (loaded
automatically with the \textbf{tidyverse} package).
\end{marginfigure}

\hypertarget{step-3.-fit-the-outcome-model}{%
\subsection{Step 3. Fit the outcome
model}\label{step-3.-fit-the-outcome-model}}

Because we are doing a weighted analysis, we can use the propensity
score, \texttt{p}, from this data frame to create our desired weights.
If we were doing a different type of analysis, we could use the
propensity score, \texttt{p}, for that purpose. We are going to add a
column to this new data frame called \texttt{weight} using the same
overlap weights used above.

\begin{Shaded}
\begin{Highlighting}[]
\NormalTok{o \textless{}{-}}\StringTok{ }\NormalTok{o }\OperatorTok{\%\textgreater{}\%}
\StringTok{  }\KeywordTok{mutate}\NormalTok{(}
    \DataTypeTok{weight =} 
      \KeywordTok{map}\NormalTok{(p, }\OperatorTok{\textasciitilde{}}\StringTok{ }\KeywordTok{case\_when}\NormalTok{(rhc}\OperatorTok{$}\NormalTok{exposure }\OperatorTok{==}\StringTok{ }\DecValTok{1} \OperatorTok{\textasciitilde{}}\StringTok{ }\DecValTok{1} \OperatorTok{{-}}\StringTok{ }\NormalTok{.x, }\OtherTok{TRUE} \OperatorTok{\textasciitilde{}}\StringTok{ }\NormalTok{.x))}
\NormalTok{  )}
\end{Highlighting}
\end{Shaded}

Now using this \texttt{weight}, we can re-run the outcome model, using
the function we created above called \texttt{run\_outcome\_mod()} along
with the \texttt{outcome\_formula} column here. We are going to create a
new column called \texttt{outcome\_model}.

\begin{Shaded}
\begin{Highlighting}[]
\NormalTok{o \textless{}{-}}\StringTok{ }\NormalTok{o }\OperatorTok{\%\textgreater{}\%}
\StringTok{  }\KeywordTok{mutate}\NormalTok{(}
    \DataTypeTok{outcome\_model =}
      \KeywordTok{map2}\NormalTok{(outcome\_formula, weight, run\_outcome\_mod)}
\NormalTok{  )}
\end{Highlighting}
\end{Shaded}

\textbf{Explore the output}

Let's take a look at what this data frame looks like now for a single
row. In the first row, we dropped \texttt{age}. First, we can examine
the propensity score formula that includes all of the variables except
\texttt{age}.

\begin{Shaded}
\begin{Highlighting}[]
\NormalTok{o[[}\StringTok{"ps\_formula"}\NormalTok{]][[}\DecValTok{1}\NormalTok{]]}
\end{Highlighting}
\end{Shaded}

\begin{verbatim}
## exposure ~ gibledhx + transhx + aps1 + wblc1 + hrt1 + pafi1 + 
##     alb1 + hema1 + bili1 + meanbp1 + paco21 + dnr1 + ph1 + resp1 + 
##     neuro + hema + sex + age + surv2md1
## <environment: 0x7fda579304c0>
\end{verbatim}

Similarly, the outcome formula includes all of the variables except
\texttt{age}.

\begin{Shaded}
\begin{Highlighting}[]
\NormalTok{o[[}\StringTok{"outcome\_formula"}\NormalTok{]][[}\DecValTok{1}\NormalTok{]] }
\end{Highlighting}
\end{Shaded}

\begin{verbatim}
## Surv(time, event) ~ exposure + gibledhx + transhx + aps1 + wblc1 + 
##     hrt1 + pafi1 + alb1 + hema1 + bili1 + meanbp1 + paco21 + 
##     dnr1 + ph1 + resp1 + neuro + hema + sex + age + surv2md1
## <environment: 0x7fda543b7f50>
\end{verbatim}

The column \texttt{p} includes the propensity scores for this updated
propensity score model. We can glance at the first six.

\begin{marginfigure}
\hypertarget{note}{%
\subsection{Note}\label{note}}

Similarly, the column \texttt{weight} includes the weights for this
updated propensity score model.
\end{marginfigure}

\begin{Shaded}
\begin{Highlighting}[]
\KeywordTok{head}\NormalTok{(o[[}\StringTok{"p"}\NormalTok{]][[}\DecValTok{1}\NormalTok{]])}
\end{Highlighting}
\end{Shaded}

\begin{verbatim}
##         1         2         3         4         5         6 
## 0.6388080 0.5367369 0.6166899 0.4688216 0.3093988 0.1441288
\end{verbatim}

Finally, the columns \texttt{ps\_model} and \texttt{outcome\_model}
include the updated model calls, using the information in the previous
columns. For example, here is our updated outcome model. We can pull the
updated coefficients out and exponentiate them.

\begin{marginfigure}
\vspace{4cm}

\hypertarget{note}{%
\subsection{Note}\label{note}}

Notice here that our result, the exposure effect, has slightly changed,
from 1.24 in the full analysis to 1.23 in this analysis that dropped
\texttt{age}.
\end{marginfigure}

\begin{Shaded}
\begin{Highlighting}[]
\NormalTok{o[[}\StringTok{"outcome\_model"}\NormalTok{]][[}\DecValTok{1}\NormalTok{]] }\OperatorTok{\%\textgreater{}\%}
\StringTok{  }\KeywordTok{coef}\NormalTok{() }\OperatorTok{\%\textgreater{}\%}
\StringTok{  }\KeywordTok{exp}\NormalTok{()}
\end{Highlighting}
\end{Shaded}

\begin{verbatim}
##   exposure   gibledhx    transhx       aps1      wblc1       hrt1      pafi1 
## 1.23530226 1.58201967 1.29875097 1.00499210 1.00059112 1.00165508 0.99967233 
##       alb1      hema1      bili1    meanbp1     paco21    dnr1Yes        ph1 
## 0.97973703 1.00178203 1.03136808 1.00070311 0.99330404 2.59442887 0.61820492 
##      resp1   neuroYes    hemaYes    sexMale        age   surv2md1 
## 0.99634972 1.40250387 1.38703246 1.07260158 1.00452756 0.07921107
\end{verbatim}

\hypertarget{step-4.-add-observed-bias-effects}{%
\subsection{Step 4. Add observed bias
effects}\label{step-4.-add-observed-bias-effects}}

We can now extract the exposure effect for each of the updated models
(using the \texttt{outcome\_model} column). This step will differ
depending on the form of your outcome model. For ours, we can use the
\texttt{coef()} and \texttt{confint()} functions to extract the
coefficients and confidence intervals.

\begin{Shaded}
\begin{Highlighting}[]
\NormalTok{o \textless{}{-}}\StringTok{ }\NormalTok{o }\OperatorTok{\%\textgreater{}\%}
\StringTok{  }\KeywordTok{mutate}\NormalTok{(}
    \DataTypeTok{point\_estimate =} 
      \KeywordTok{map\_dbl}\NormalTok{(outcome\_model, }\OperatorTok{\textasciitilde{}}\StringTok{ }\KeywordTok{exp}\NormalTok{(}\KeywordTok{coef}\NormalTok{(.x)[}\StringTok{"exposure"}\NormalTok{])),}
    \DataTypeTok{lb =} 
      \KeywordTok{map\_dbl}\NormalTok{(outcome\_model, }\OperatorTok{\textasciitilde{}}\StringTok{ }\KeywordTok{exp}\NormalTok{(}\KeywordTok{confint}\NormalTok{(.x, }\StringTok{"exposure"}\NormalTok{)[, }\DecValTok{1}\NormalTok{])),}
    \DataTypeTok{ub =} 
      \KeywordTok{map\_dbl}\NormalTok{(outcome\_model, }\OperatorTok{\textasciitilde{}}\StringTok{ }\KeywordTok{exp}\NormalTok{(}\KeywordTok{confint}\NormalTok{(.x, }\StringTok{"exposure"}\NormalTok{)[, }\DecValTok{2}\NormalTok{]))}
\NormalTok{  )}
\end{Highlighting}
\end{Shaded}

Now we have the observed bias effects, defined as the updated hazard
ratio, along with a lower bound and upper bound for each the analyses
dropping each of the indicated variables.

\begin{Shaded}
\begin{Highlighting}[]
\NormalTok{o }\OperatorTok{\%\textgreater{}\%}
\StringTok{  }\KeywordTok{select}\NormalTok{(dropped, point\_estimate, lb, ub)}
\end{Highlighting}
\end{Shaded}

\begin{verbatim}
## # A tibble: 24 x 4
##    dropped  point_estimate    lb    ub
##    <chr>             <dbl> <dbl> <dbl>
##  1 renalhx            1.24  1.11  1.37
##  2 gibledhx           1.23  1.10  1.36
##  3 transhx            1.25  1.12  1.38
##  4 aps1               1.25  1.12  1.38
##  5 wblc1              1.24  1.11  1.37
##  6 hrt1               1.25  1.13  1.39
##  7 pafi1              1.25  1.13  1.38
##  8 alb1               1.24  1.11  1.37
##  9 hema1              1.23  1.11  1.37
## 10 bili1              1.24  1.11  1.37
## # ... with 14 more rows
\end{verbatim}

\hypertarget{step-5.-add-tipping-point}{%
\subsection{Step 5. Add tipping point}\label{step-5.-add-tipping-point}}

In addition to observing the effect of dropping each variable and groups
of variables, we can also add rows that demonstrate what the effect
would look like if it were shifted to be null, in this case crossing 1.
To do this we use the \texttt{observed\_bias\_tip()} function from the
\textbf{tipr} package.

For this analysis, we are going to do this twice, once to see how the
effect would look if the lower bound crossed 1 (plugging in
\texttt{full\_model\$lb} for the \texttt{tip} argument), and once to see
how the effect would look if the point estimate crossed 1 (plugging in
\texttt{full\_model\$point\_estimate} for the \texttt{tip} argument).

\begin{marginfigure}
\hypertarget{tipr-function}{%
\subsection{\texorpdfstring{\textbf{tipr}
function}{tipr function}}\label{tipr-function}}

The \texttt{observed\_bias\_tip()} function takes five arguments:

\begin{itemize}
\tightlist
\item
  \texttt{tip}: The value you would like to shift
\item
  \texttt{point\_estimate}: The resulting effect from the full model (in
  our case the hazard ratio we've saved in the \texttt{full\_model}
  object)
\item
  \texttt{lb}: The resulting lower bound from the full model
\item
  \texttt{ub}: The resulting upper bound from the full model
\item
  \texttt{tip\_desc}: A description of the tipping point you've chosen
\end{itemize}
\end{marginfigure}

\begin{Shaded}
\begin{Highlighting}[]
\NormalTok{o \textless{}{-}}\StringTok{ }\KeywordTok{bind\_rows}\NormalTok{(}
\NormalTok{  o,}
  \KeywordTok{observed\_bias\_tip}\NormalTok{(}
    \DataTypeTok{tip =}\NormalTok{ full\_model}\OperatorTok{$}\NormalTok{lb,}
\NormalTok{    full\_model}\OperatorTok{$}\NormalTok{point\_estimate,}
\NormalTok{    full\_model}\OperatorTok{$}\NormalTok{lb,}
\NormalTok{    full\_model}\OperatorTok{$}\NormalTok{ub,}
    \StringTok{"Hypothetical unmeasured confounder (Tip LB)"}\NormalTok{),}
  \KeywordTok{observed\_bias\_tip}\NormalTok{(}
    \DataTypeTok{tip =}\NormalTok{ full\_model}\OperatorTok{$}\NormalTok{point\_estimate,}
\NormalTok{    full\_model}\OperatorTok{$}\NormalTok{point\_estimate,}
\NormalTok{    full\_model}\OperatorTok{$}\NormalTok{lb,}
\NormalTok{    full\_model}\OperatorTok{$}\NormalTok{ub,}
    \StringTok{"Hypothetical unmeasured confounder (Tip Point Est)"}\NormalTok{)}
\NormalTok{)}
\end{Highlighting}
\end{Shaded}

\hypertarget{step-6.-add-observed-covariate-e-value}{%
\subsection{Step 6. Add Observed Covariate
E-value}\label{step-6.-add-observed-covariate-e-value}}

\begin{marginfigure}
\hypertarget{tipr-function}{%
\subsection{\texorpdfstring{\textbf{tipr}
function}{tipr function}}\label{tipr-function}}

The \texttt{observed\_covariate\_e\_value()} function takes five
arguments:

\begin{itemize}
\tightlist
\item
  \texttt{lb}: The lower bound from the full model (in our case
  \texttt{full\_model\$lb})
\item
  \texttt{ub}: The upper bound from the full model
\item
  \texttt{lb\_adj}: The lower bound from the adjusted model (in our case
  \texttt{o\$lb})
\item
  \texttt{ub\_adj}: The upper bound from the adjusted model
\item
  \texttt{transform}: The transformation (if any) you would like to
  perform if the input is not a risk ratio, based on VanderWeele and
  Ding's suggestions. In our case, we will input ``HR''.
\end{itemize}

Notice here we are using the \texttt{map2\_dbl()} function so that we
can iterate over every column in our observed bias data frame. If we
wanted to calculate the Observed Covariate E-value for a single model,
we could just use the \texttt{observed\_covariate\_e\_value()} function
alone. For example, using the variable \texttt{dnr1}, we would plug in
the following.

\begin{verbatim}
observed_covariate_e_value(
  lb = 1.11,
  ub = 1.37,
  lb_adj = 1.00,
  ub_adj = 1.23, 
  transform = "HR")
## [1] 1.358969
\end{verbatim}
\end{marginfigure}

Let's add a column that creates the Observed Covariate E-value. To do
this, we will use the \texttt{observed\_covariate\_e\_value()} function
from the \textbf{tipr} package. Since we have hazard ratios, we can use
the transformation suggested by \citet{VanderWeele:2017ki}.

\begin{Shaded}
\begin{Highlighting}[]
\NormalTok{o \textless{}{-}}\StringTok{ }\NormalTok{o }\OperatorTok{\%\textgreater{}\%}
\StringTok{  }\KeywordTok{mutate}\NormalTok{(}
    \DataTypeTok{e\_value =} \KeywordTok{map2\_dbl}\NormalTok{(}
\NormalTok{      lb,}
\NormalTok{      ub,}
      \OperatorTok{\textasciitilde{}}\StringTok{ }\KeywordTok{observed\_covariate\_e\_value}\NormalTok{(}
        \DataTypeTok{lb =}\NormalTok{ full\_model}\OperatorTok{$}\NormalTok{lb,}
        \DataTypeTok{ub =}\NormalTok{ full\_model}\OperatorTok{$}\NormalTok{ub,}
        \DataTypeTok{lb\_adj =}\NormalTok{ .x,}
        \DataTypeTok{ub\_adj =}\NormalTok{ .y,}
        \DataTypeTok{transform =} \StringTok{"HR"}\NormalTok{)}
\NormalTok{    )}
\NormalTok{  )}
\end{Highlighting}
\end{Shaded}

\hypertarget{step-7.-create-the-observed-bias-plot}{%
\subsection{Step 7. Create the Observed bias
plot}\label{step-7.-create-the-observed-bias-plot}}

We have calculated our observed bias effects and Observed Covariate
E-values, now we are going to create a publication-ready plot that
displays these. There are a few steps we can take to improve the
aesthetics of the plot.

\begin{itemize}
\tightlist
\item
  Update variable names with the appropriate variable labels.
\item
  Re-order the data frame so it will display the plot in a logical
  manner. For example, we could order the data frame, \texttt{o}, by the
  lower bound of the observed bias effect, in our case \texttt{lb}.
\end{itemize}

We have created a small function, \texttt{observed\_bias\_order()} in
\textbf{tipr} to assist with ordering your observed bias data frame for
the observed bias plot. This will output the same data frame in the
correct order.

\begin{Shaded}
\begin{Highlighting}[]
\NormalTok{o \textless{}{-}}\StringTok{ }\KeywordTok{observed\_bias\_order}\NormalTok{(o, }\StringTok{"lb"}\NormalTok{)}
\end{Highlighting}
\end{Shaded}

\begin{marginfigure}
\hypertarget{tipr-function}{%
\subsection{\texorpdfstring{\textbf{tipr}
function}{tipr function}}\label{tipr-function}}

The \texttt{observed\_bias\_order()} function takes two arguments:

\begin{itemize}
\tightlist
\item
  \texttt{d}: Your observed bias data frame (in our case, \texttt{o})
\item
  \texttt{by}: The name of the variable to order by (in our case we are
  going to order by the lower bound of the adjusted effect, \texttt{lb})
\end{itemize}
\end{marginfigure}

Now we're ready to make the plot.

\begin{marginfigure}
\hypertarget{note}{%
\subsection{Note}\label{note}}

Before creating the plot, we like to merge in variable labels. To do
this, we create a file, \texttt{var\_labels.csv}, with two columns: the
variable name and the label, for example

\begin{verbatim}
var,var_label
adld3p,ADL
age,Age
...
\end{verbatim}

We then read this into R and left join it with our data frame,
\texttt{o}.

\begin{verbatim}
o <- o %>%
  left_join(
    read_csv("var_labels.csv"),
    by = c("dropped" = "var")
  ) %>%
  mutate(
    dropped = case_when(
      type == "covariate" ~ var_label,
      TRUE ~ dropped
    ),
    dropped = glue::glue("Dropped {dropped}")
  )
\end{verbatim}
\end{marginfigure}

We are going to demonstrate how this is creating by building the plot up
in layers. Let's begin with the \texttt{ggplot()} function, giving it
our observed bias data frame, \texttt{o}.

\begin{Shaded}
\begin{Highlighting}[]
\NormalTok{g \textless{}{-}}\StringTok{ }\KeywordTok{ggplot}\NormalTok{(}\DataTypeTok{data =}\NormalTok{ o)}
\end{Highlighting}
\end{Shaded}

In order to demonstrate what our original result was with the full
model, we can add a line at the full model's point estimate along with a
rectangle spanning the confidence bounds. In addition, we add a dashed
black line to indicate the null value, 1.

\begin{Shaded}
\begin{Highlighting}[]
\NormalTok{g \textless{}{-}}\StringTok{ }\NormalTok{g }\OperatorTok{+}\StringTok{ }
\StringTok{  }\KeywordTok{geom\_hline}\NormalTok{(}\DataTypeTok{yintercept =} \KeywordTok{c}\NormalTok{(full\_model}\OperatorTok{$}\NormalTok{lb,}
\NormalTok{                            full\_model}\OperatorTok{$}\NormalTok{point\_estimate,}
\NormalTok{                            full\_model}\OperatorTok{$}\NormalTok{ub),}
             \DataTypeTok{lwd =} \KeywordTok{c}\NormalTok{(}\DecValTok{1}\NormalTok{, }\DecValTok{2}\NormalTok{, }\DecValTok{1}\NormalTok{),}
             \DataTypeTok{lty =} \KeywordTok{c}\NormalTok{(}\DecValTok{2}\NormalTok{, }\DecValTok{1}\NormalTok{, }\DecValTok{2}\NormalTok{),}
             \DataTypeTok{color =} \StringTok{"light blue"}\NormalTok{) }\OperatorTok{+}\StringTok{ }
\StringTok{  }\KeywordTok{geom\_rect}\NormalTok{(}\DataTypeTok{ymin =}\NormalTok{ full\_model}\OperatorTok{$}\NormalTok{lb,}
            \DataTypeTok{ymax =}\NormalTok{ full\_model}\OperatorTok{$}\NormalTok{ub,}
            \DataTypeTok{xmin =} \DecValTok{0}\NormalTok{,}
            \DataTypeTok{xmax =} \DecValTok{27}\NormalTok{,}
            \DataTypeTok{alpha =} \FloatTok{0.01}\NormalTok{,}
            \DataTypeTok{fill =} \StringTok{"light blue"}\NormalTok{) }\OperatorTok{+}
\StringTok{  }\KeywordTok{geom\_hline}\NormalTok{(}\DataTypeTok{yintercept =} \DecValTok{1}\NormalTok{, }\DataTypeTok{lty =} \DecValTok{2}\NormalTok{)}
\end{Highlighting}
\end{Shaded}

Now we are going to add the observed bias effects (the point estimate,
lower bound, and upper bound of each of the models fit after dropping
the variable(s)). We do this by feeding the data frame \texttt{o} to the
\texttt{ggplot()} function and using the function
\texttt{geom\_pointrange()} to create a point at the point estimate
along with a line for the width of the confidence interval. We are going
to also use the \texttt{cord\_flip()} function to flip the coordinates
so that the plot is oriented vertically rather than horizontally.

\begin{Shaded}
\begin{Highlighting}[]
\NormalTok{g \textless{}{-}}\StringTok{ }\NormalTok{g }\OperatorTok{+}\StringTok{ }
\StringTok{  }\KeywordTok{geom\_pointrange}\NormalTok{(}\KeywordTok{aes}\NormalTok{(}\DataTypeTok{x =}\NormalTok{ dropped,}
                      \DataTypeTok{y =}\NormalTok{ point\_estimate,}
                      \DataTypeTok{ymin =}\NormalTok{ lb,}
                      \DataTypeTok{ymax =}\NormalTok{ ub)) }\OperatorTok{+}\StringTok{ }
\StringTok{  }\KeywordTok{coord\_flip}\NormalTok{()}
\end{Highlighting}
\end{Shaded}

\begin{marginfigure}
\hypertarget{note}{%
\subsection{Note}\label{note}}

Here we are demonstrating how to build the observed bias plots using the
\texttt{ggplot()} function from the \textbf{ggplot2} package (loaded
automatically with the \textbf{tidyverse} package). The tools
demonstrated here from the \textbf{tipr} package can work with any
plotting functions.
\end{marginfigure}

Let's add some aesthetic updates, such as x- and y-axis labels, a theme,
etc.

\begin{Shaded}
\begin{Highlighting}[]
\NormalTok{g \textless{}{-}}\StringTok{ }\NormalTok{g }\OperatorTok{+}\StringTok{ }
\StringTok{  }\KeywordTok{ylab}\NormalTok{(}\StringTok{"Effect modeled without variable(s) of interest"}\NormalTok{) }\OperatorTok{+}
\StringTok{  }\KeywordTok{xlab}\NormalTok{(}\StringTok{""}\NormalTok{) }\OperatorTok{+}
\StringTok{  }\KeywordTok{ggtitle}\NormalTok{(}\StringTok{"A. Most influential covariates"}\NormalTok{) }\OperatorTok{+}
\StringTok{  }\KeywordTok{theme\_bw}\NormalTok{()}
\end{Highlighting}
\end{Shaded}

Next, we can create a plot of the Observed Covariate E-values as well as
the traditional E-values.

\begin{Shaded}
\begin{Highlighting}[]
\NormalTok{e \textless{}{-}}\StringTok{ }\KeywordTok{ggplot}\NormalTok{(}\DataTypeTok{data =}\NormalTok{ o) }\OperatorTok{+}\StringTok{ }
\StringTok{  }\KeywordTok{geom\_point}\NormalTok{(}\KeywordTok{aes}\NormalTok{(}\DataTypeTok{x =}\NormalTok{ dropped, }\DataTypeTok{y =}\NormalTok{ e\_value, }\DataTypeTok{color =}\NormalTok{ type),}
             \DataTypeTok{pch =} \DecValTok{8}\NormalTok{) }\OperatorTok{+}\StringTok{ }
\StringTok{  }\KeywordTok{scale\_color\_manual}\NormalTok{(}\DataTypeTok{name =} \StringTok{""}\NormalTok{,}
                     \DataTypeTok{values =} \KeywordTok{c}\NormalTok{(}\StringTok{"covariate"}\NormalTok{ =}\StringTok{ "purple"}\NormalTok{,}
                                \StringTok{"group"}\NormalTok{ =}\StringTok{ "orange"}\NormalTok{,}
                                \StringTok{"tip"}\NormalTok{ =}\StringTok{ "red"}\NormalTok{),}
                     \DataTypeTok{labels =} \KeywordTok{c}\NormalTok{(}\StringTok{"Observed Covariate E{-}value"}\NormalTok{,}
                                \StringTok{"Observed Covariate E{-}value (group)"}\NormalTok{,}
                                \StringTok{"E{-}value"}\NormalTok{)}
\NormalTok{  ) }\OperatorTok{+}\StringTok{ }
\StringTok{  }\KeywordTok{coord\_flip}\NormalTok{()}
\end{Highlighting}
\end{Shaded}

We can add some aesthetic updates, such as x- and y-axis labels, a
theme, etc to the E-values plot.

\begin{Shaded}
\begin{Highlighting}[]
\NormalTok{e \textless{}{-}}\StringTok{ }\NormalTok{e }\OperatorTok{+}\StringTok{ }
\StringTok{  }\KeywordTok{ylab}\NormalTok{(}\StringTok{"OCE"}\NormalTok{) }\OperatorTok{+}
\StringTok{  }\KeywordTok{scale\_x\_discrete}\NormalTok{(}\DataTypeTok{name =} \StringTok{""}\NormalTok{, }\DataTypeTok{labels =} \KeywordTok{element\_blank}\NormalTok{()) }\OperatorTok{+}
\StringTok{  }\KeywordTok{theme\_bw}\NormalTok{() }\OperatorTok{+}
\StringTok{  }\KeywordTok{ggtitle}\NormalTok{(}\StringTok{"B. Sensitivity to unmeasured confounding"}\NormalTok{) }\OperatorTok{+}
\StringTok{  }\KeywordTok{theme}\NormalTok{(}\DataTypeTok{legend.title =} \KeywordTok{element\_blank}\NormalTok{())}
\end{Highlighting}
\end{Shaded}

We can use the \texttt{grid.arrange()} function to combine the two
plots. Let's see how that looks!

\begin{Shaded}
\begin{Highlighting}[]
\KeywordTok{grid.arrange}\NormalTok{(g, e, }\DataTypeTok{widths =} \KeywordTok{c}\NormalTok{(}\DecValTok{2}\NormalTok{, }\DecValTok{1}\NormalTok{), }\DataTypeTok{ncol =} \DecValTok{2}\NormalTok{)}
\end{Highlighting}
\end{Shaded}

\begin{figure*}[h]
\includegraphics{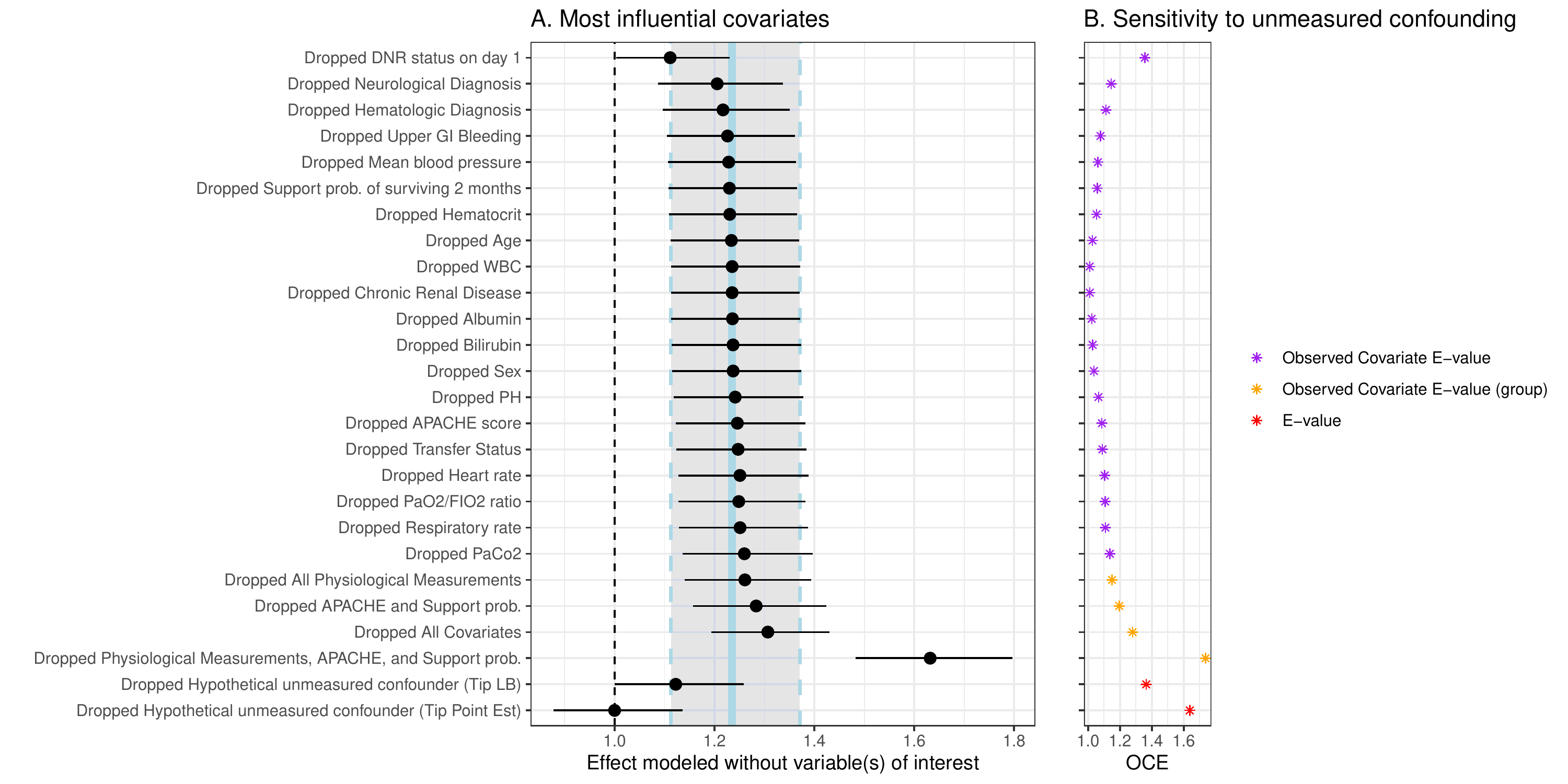} \end{figure*}

We are going to examine how removing each variable, one at a time, will
affect the final result as well as how removing two groups of variables,
``Labs'' and ``Physiological Measurements'', will affect the result.

\hypertarget{choose-variable-groups-to-drop-simultaneously}{%
\subsection{Choose variable groups to drop
simultaneously}\label{choose-variable-groups-to-drop-simultaneously}}

We will first create a named list of the two groups of variables.

\begin{Shaded}
\begin{Highlighting}[]
\NormalTok{groups \textless{}{-}}\StringTok{ }\KeywordTok{list}\NormalTok{(}
  \StringTok{"No Labs"}\NormalTok{ =}\StringTok{ }
\StringTok{    }\KeywordTok{c}\NormalTok{(}\StringTok{"wblc1"}\NormalTok{, }\StringTok{"pafi1"}\NormalTok{, }\StringTok{"alb1"}\NormalTok{, }\StringTok{"hema1"}\NormalTok{, }\StringTok{"bili1"}\NormalTok{, }\StringTok{"paco21"}\NormalTok{, }\StringTok{"ph1"}\NormalTok{),}
  \StringTok{"No Physiological Measurements"}\NormalTok{ =}\StringTok{ }
\StringTok{    }\KeywordTok{c}\NormalTok{(}\StringTok{"aps1"}\NormalTok{, }\StringTok{"hrt1"}\NormalTok{, }\StringTok{"wblc1"}\NormalTok{, }\StringTok{"pafi1"}\NormalTok{, }\StringTok{"alb1"}\NormalTok{, }\StringTok{"hema1"}\NormalTok{, }\StringTok{"bili1"}\NormalTok{,}
      \StringTok{"paco21"}\NormalTok{, }\StringTok{"meanbp1"}\NormalTok{, }\StringTok{"resp1"}\NormalTok{, }\StringTok{"ph1"}\NormalTok{))}
\end{Highlighting}
\end{Shaded}

\newpage
\newpage

\bibliography{citations.bib}